\documentclass[aps,prl,reprint,groupedaddress]{revtex4-1}

\usepackage{graphicx}
\usepackage{amsmath}

\begin{document}

\title{A universal 3D imaging sensor on a silicon photonics platform}

\author{Christopher Rogers$^{1,\ast}$}
\author{Alexander Y. Piggott$^{1,\ast}$}
\author{David J. Thomson$^2$}
\author{Robert F. Wiser$^1$}
\author{Ion E. Opris$^3$}
\author{Steven A. Fortune$^1$}
\author{Andrew J. Compston$^1$}
\author{Alexander Gondarenko$^1$}
\author{Fanfan Meng$^2$}
\author{Xia Chen$^2$}
\author{Graham T. Reed$^2$}
\author{Remus Nicolaescu$^{1,\dagger}$}
\date{July 27, 2020}

\affiliation{$^1$ Pointcloud Inc, San Francisco, California 94107, USA}
\affiliation{$^2$ Optoelectronics Research Centre, University of Southampton, University Road Southampton, Hampshire SO17 1BJ, UK}
\affiliation{Opris Consulting, San Jose, California 95138, USA}
\affiliation{$\ast$ These authors contributed equally to the work.}
\affiliation{$\dagger$ Corresponding author: remus.nicolaescu@point.cloud}

\begin{abstract}
Accurate 3D imaging is essential for machines to map and interact with the physical world\cite{curmson_jofr2008, qwang_aei2019}. While numerous 3D imaging technologies exist, each addressing niche applications with varying degrees of success, none have achieved the breadth of applicability and impact that digital image sensors have achieved in the 2D imaging world\cite{ddlichti_ijprs2007, jkidd_mscthesis2017, jsalvi_pr2004, acorti_ras2016, pmcmanamon_oe2012, pfmcmanamon_oe2017, swhutchings_ijssc2019, arximenes_ijscc2019}. A large-scale two-dimensional array of coherent detector pixels operating as a light detection and ranging (LiDAR) system could serve as a universal 3D imaging platform. Such a system would offer high depth accuracy and immunity to interference from sunlight, as well as the ability to directly measure the velocity of moving objects\cite{bbehroozpour_ieeecm2017}. However, due to difficulties in providing electrical and photonic connections to every pixel, previous systems have been restricted to fewer than 20 pixels\cite{faflatouni_oe2015, amartin_jlt2018, dinoue_oe2019, cli_oe2019}. Here, we demonstrate the first large-scale coherent detector array consisting of 512 ($32 \times 16$) pixels, and its operation in a 3D imaging system. Leveraging recent advances in the monolithic integration of photonic and electronic circuits, a dense array of optical heterodyne detectors is combined with an integrated electronic readout architecture, enabling straightforward scaling to arbitrarily large arrays. Meanwhile, two-axis solid-state beam steering eliminates any tradeoff between field of view and range. Operating at the quantum noise limit\cite{mjcollett_jmo1987, marubin_ol2007}, our system achieves an accuracy of $3.1~\mathrm{mm}$ at a distance of 75 metres using only $4~\mathrm{mW}$ of light, an order of magnitude more accurate than existing solid-state systems at such ranges. Future reductions of pixel size using state-of-the-art components could yield resolutions in excess of 20 megapixels for arrays the size of a consumer camera sensor. This result paves the way for the development and proliferation of low cost, compact, and high performance 3D imaging cameras, enabling new applications from robotics and autonomous navigation to augmented reality and healthcare.

\begin{center}
  \emph{This paper is dedicated to the memory of Sunil Sandhu.}
\end{center}
\end{abstract}

\maketitle

The digital complementary metal--oxide--semiconductor (CMOS) image sensor revolutionized 2D imaging, borrowing technology from silicon microelectronics to produce a flexible and scalable camera sensor\cite{aelgamal_ieeecdm2005}. As a focal plane array (FPA), the digital image sensor uses a lens to focus light and form an image on the detector. A key advantage of this scheme is that the field of view and light collection efficiency are not set by the image sensor, but by the choice of lens. Furthermore, the CMOS image sensor can be optimized for high performance or low-cost. The flexibility of this arrangement made the digital CMOS sensor the sensor of choice for most 2D imaging.

In contrast, 3D imaging is characterized by an assortment of competing technologies, each addressing a narrow niche. Long range and high precision applications such as autonomous vehicles and construction site mapping are dominated by expensive and fragile mechanically steered LiDAR systems\cite{ddlichti_ijprs2007, jkidd_mscthesis2017}. Solid-state solutions such as structured light\cite{jsalvi_pr2004} and time-of-flight arrays\cite{acorti_ras2016, pmcmanamon_oe2012, pfmcmanamon_oe2017, swhutchings_ijssc2019, arximenes_ijscc2019, blstann_lrta2004, khu_ieeesj2017} are used when affordability, compactness, and reliability must be achieved at the expense of performance, such as in mobile devices and augmented reality systems. Optical phased arrays are a promising solid-state approach, but developing long-range 2D-scanning systems has proven challenging, with current demonstrations limited to less than 20 metres\cite{cvpoulton_ol2017, samiller_cleo2018, cvpoulton_ieeeqe2019}. As such, no current technology addresses the needs of these diverse use cases.

Here, we demonstrate a fully solid-state, integrated photonic LiDAR based on the same FPA concept as the CMOS image sensor. By efficient use of light, our system achieves the range, depth accuracy and field of view needed by demanding applications such as self driving vehicles\cite{curmson_jofr2008} and drone- or land-based 3D mapping\cite{jwang_jirs2015, akasturi_lrta2016}. The centerpiece of our system is the coherent receiver array, a focal plane array of compact optical heterodyne detectors operating at the quantum noise limit\cite{mjcollett_jmo1987, marubin_ol2007}. To eliminate tradeoffs between field of view and range, the receiver is paired with solid-state beam steering that sequentially illuminates the scene in small patches. The coherent receiver array allows our architecture to operate using a frequency-modulated continuous-wave (FMCW) coherent LiDAR scheme\cite{hdgriffiths_ecej1990, jriemensberger_nature2020}. In contrast to time-of-flight LiDARs that use pulses of light, FMCW LiDAR uses a linearly chirped laser. Scattered light received from the target is mixed with local oscillator light in a heterodyne receiver, producing a beat frequency proportional to round-trip travel time and hence distance to the target.

The FMCW scheme confers several advantages relative to time-of-flight schemes. First, heterodyne detection is immune to interference from sunlight and other nearby LiDAR systems since it selectively detects light close in frequency to the local oscillator light\cite{bbehroozpour_ieeecm2017}. Second, coherent LiDAR directly measures target velocity through Doppler shifts of the received light\cite{hdgriffiths_ecej1990, jriemensberger_nature2020}. Third, high depth accuracy, depending upon only chirp bandwidth and signal-to-noise ratio\cite{kthurn_ieeemtt2013}, is achieved using relatively low frequency receiver electronics. In contrast, depth accuracy is limited by receiver bandwidth for time-of-flight schemes. Finally, the FMCW system emits photons continuously and thus is well suited for photonic integration, where non-linear effects constrain peak power\cite{hktsang_apl2002, hrong_nature2005}. Conversely, time-of-flight schemes emit photons in short high-power bursts.

Previous 3D imaging systems based on coherent receiver arrays have been limited to fewer than 20 pixels due to their reliance on direct electrical connections to each pixel\cite{faflatouni_oe2015, amartin_jlt2018, dinoue_oe2019, cli_oe2019}. To address scalability, we implemented our LiDAR system on a silicon photonics process with monolithically integrated radio-frequency (RF) CMOS electronics\cite{kgiewont_ieeeqe2019}. A highly multiplexed electronic readout architecture is integrated into the receiver array, minimizing external electrical connections while maintaining signal integrity. Our 512-pixel prototype array can thus be scaled to arbitrarily large numbers of pixels by increasing the size of the array. Furthermore, the use of a standard commercial foundry process facilitates immediate mass production at minimal cost.

\begin{figure*}[t]
\centering
\includegraphics[width=0.85\textwidth]{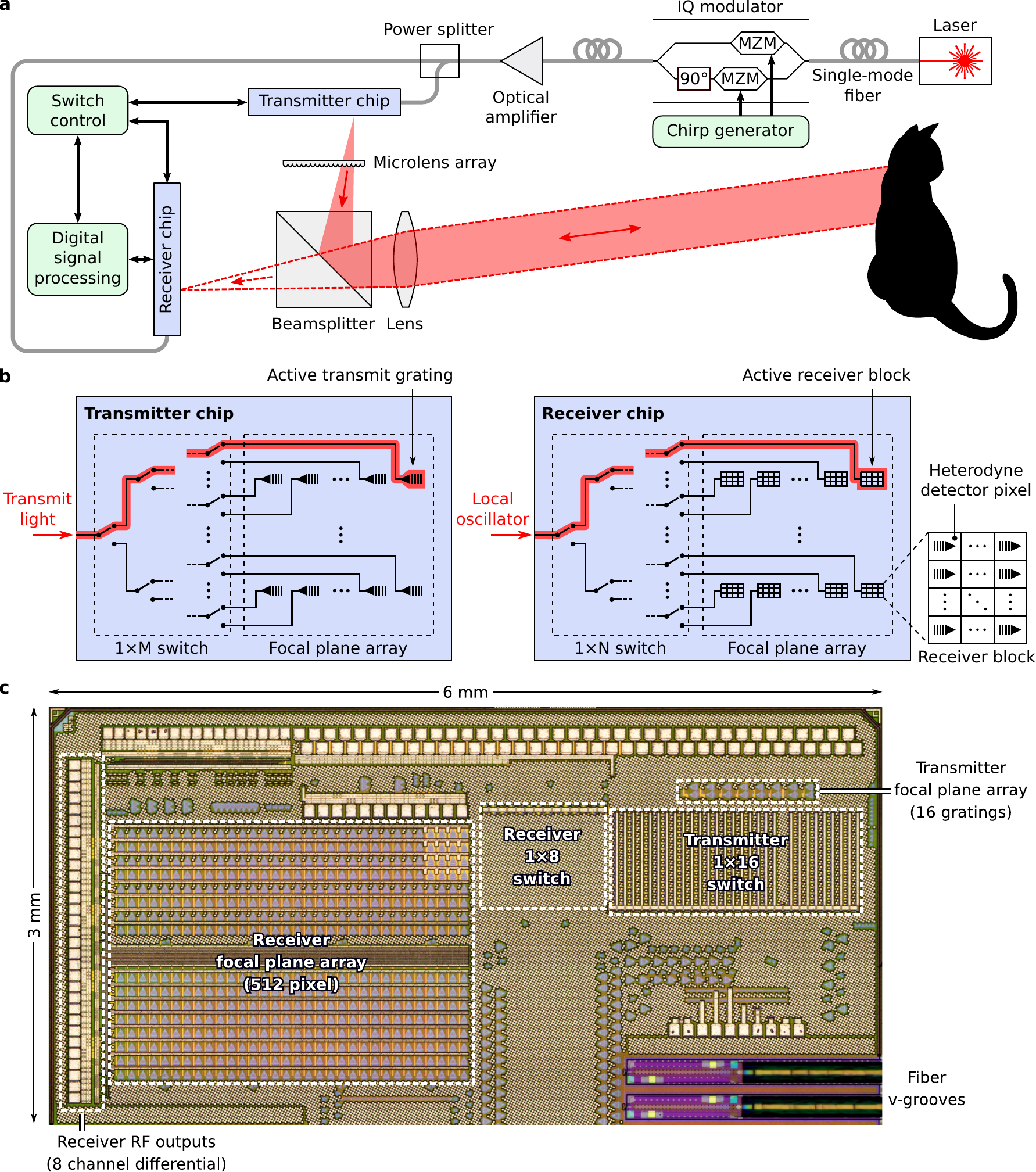}
\caption{Solid-state 3D imaging architecture. (a) Our architecture consists of two focal plane arrays (FPAs): a transmitter FPA that sequentially illuminates patches of the scene, and a receiver FPA that detects scattered light from the scene. The frequency-modulated continuous-wave scheme is used for ranging. An optional microlens array can be used to shape the illumination pattern to more closely match the receiver array, thereby improving system efficiency. (b) On-chip steering of light is provided by thermo-optic switching trees on both the transmitter and receiver chips. (c) Optical micrograph of our demonstrator chip, showing the switching trees and focal plane arrays for both the transmit and receive functionality.}
\label{fig:1_architecture}
\end{figure*}

\section*{Scalable 3D imaging architecture}
As shown in Fig. \ref{fig:1_architecture}(a), our architecture is based on two FPAs. The first acts as a transmitter, and the second as a receiver. Chirped laser light for the FMCW scheme is generated externally by modulating a fixed-frequency $1550~\mathrm{nm}$ laser with a silicon-photonic IQ Mach-Zehnder modulator (MZM), which is in turn driven by an arbitrary waveform generator. This approach ensures chirp linearity and enables the use of a simple, low-noise laser.

Long-range performance is achieved by sequentially illuminating and reading out the scene in small patches. By illuminating only the portion of the scene corresponding to the pixels being read out, no illumination light is wasted. As illustrated in Fig. \ref{fig:1_architecture}(b), the transmitter consists of a switching tree terminated by a FPA of grating couplers. Light is directed to one transmit grating at a time, illuminating a small subset of the scene. This switching approach to beam steering is robust and can be scaled up to arbitrarily large arrays, with optical losses limited only by waveguide scattering\cite{yshen_nphoton2017} and the extinction ratio of the switching trees. Meanwhile, the receiver consists of a dense FPA of miniaturized heterodyne receivers. All receiver pixels that correspond to the illuminated area are simultaneously read out in parallel. Since the angular resolution is defined by the point spread function of the lens, which drops off very quickly, there is negligible crosstalk between different receiver pixels. To avoid wasting local oscillator light, a second switching tree provides only the activated subset of the receiver FPA with local oscillator light. 

Parallel readout in the receiver is fundamental to this architecture’s scalability. First, resolution is defined by the number of pixels in the receiver FPA, rather than the number of steering positions. This increases the maximum resolution for a given chip size since heterodyne receiver pixels are $\sim 10 \times$ smaller than thermo-optic switches. Second, fast thermo-optic switching is unnecessary because the pixel rate is decoupled from the switching rate. Finally, parallel readout proportionally reduces the receiver signal frequencies for an FMCW scheme by allowing longer ramp times, simplifying the readout electronics.

\section*{Design of hybrid CMOS-photonics chip}
\begin{figure*}[t]
\centering
\includegraphics[width=\textwidth]{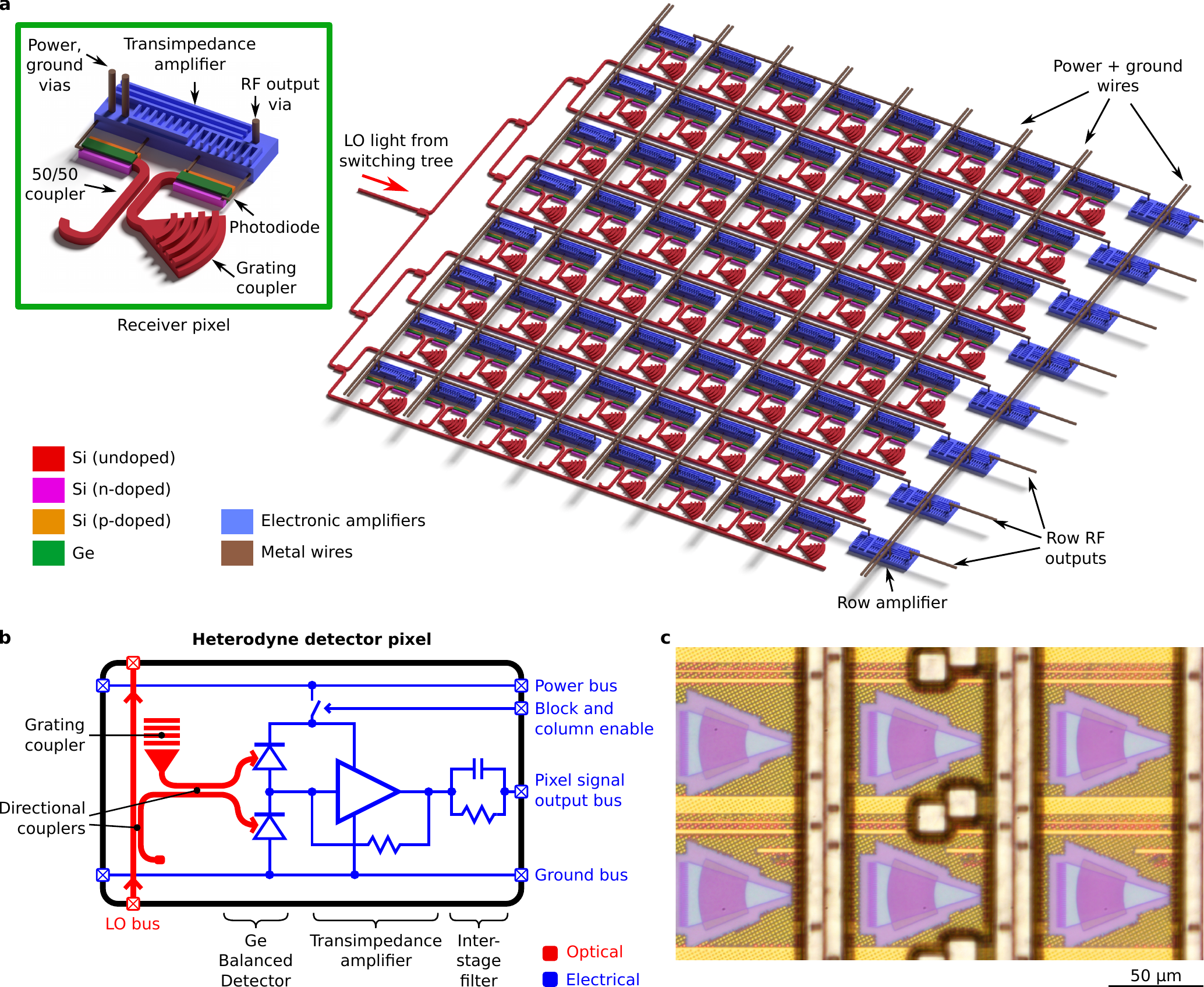}
\caption{Receiver focal plane array design. (a) Schematic of a receiver block in our receiver focal plane array. Within the receiver block, local oscillator (LO) light is distributed to a dense array of heterodyne detector pixels via a network of silicon waveguides. Meanwhile, each pixel collects scattered light from the scene using a grating coupler, which is combined with LO light on a balanced detector to produce a detectable photocurrent. The photocurrent is amplified in two stages: first by a transimpedance amplifier within the pixel, and again by an amplifier at the end of each row. For clarity, we have omitted control wires from the diagram. (b) Electrical schematic of the heterodyne detector pixel. (c) Optical micrograph of a small subset of the receiver focal plane array.}
\label{fig:2_coherent_fpa}
\end{figure*}

An optical micrograph of our demonstrator chip is shown in Fig. \ref{fig:1_architecture}(c). The transmitter consists of a $1 \times 16$ thermo-optic switch tree with 16 grating couplers in the transmit FPA. Meanwhile, the receiver consists of a $32 \times 16$ (512) pixel array of heterodyne receivers, with local oscillator light provided by a $1 \times 8$ switch tree. A subsection of the receiver is schematically illustrated in Fig. \ref{fig:2_coherent_fpa}(a). Each pixel collects scattered light from the scene using a grating coupler. Local oscillator light is provided to each pixel via a network of silicon waveguides. Scattered light and local oscillator (LO) are mixed on a balanced detector consisting of a 50-50 directional coupler and germanium PIN photodiodes, producing a heterodyne tone corresponding to the target's distance. The signal is then amplified by a transimpedance amplifier (TIA) integrated within the pixel. A buffer amplifier at the end of each row transmits the signal to the edge of the chip, maintaining wide bandwidths while driving the large parasitic capacitances of the wiring and multiplexed circuitry. Finally, a set of 8 output amplifiers transmits the signal off-chip, enabling parallel readout. As shown in Fig. \ref{fig:2_coherent_fpa}(b), the individual pixels are turned on and off using a power switch built into each TIA.

\begin{figure*}[t]
\centering
\includegraphics[width=0.813\textwidth]{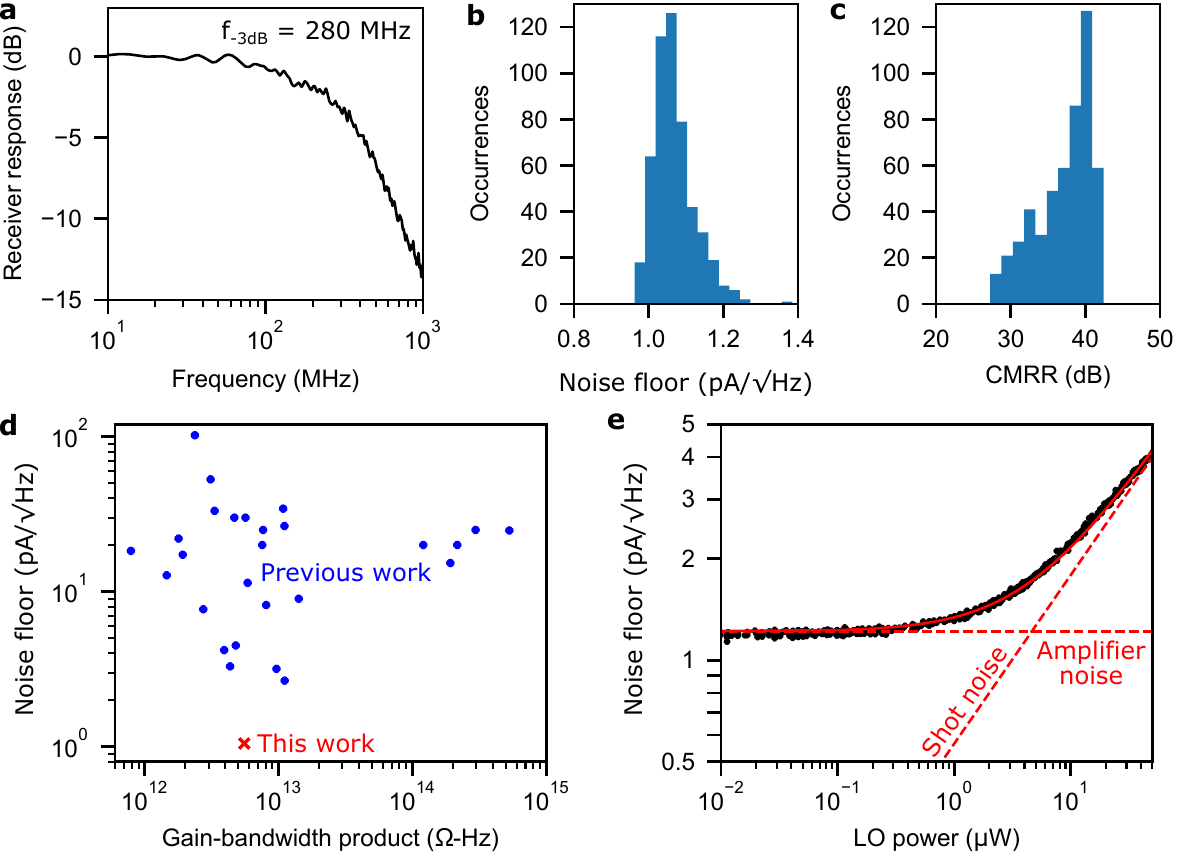}
\caption{Receiver electro-optic performance. (a) Measured frequency response of the receiver readout chain for an optical signal supplied to a single pixel, showing a cutoff frequency of $280~\mathrm{MHz}$. (b, c) Histograms of input-referred amplifier noise and common-mode rejection ratio (CMRR) respectively throughout the full array, showing tight distributions for both parameters. (d) Largely due to tight integration between our photodiodes and transimpedance amplifiers, we have achieved a high gain-bandwidth product with significantly improved noise performance compared to previous designs. (e) Input-referred noise as a function of optical local oscillator (LO) power for a single pixel, demonstrating shot-noise limited detection using $<10~\mathrm{\mu W}$ of LO power.
}
\label{fig:3_electrooptic_perf}
\end{figure*}

In general, minimizing the input-referred noise of the electronic signal chain improves the pixel's sensitivity and detection probability. Furthermore, higher receiver bandwidths are desirable since this increases the maximum range for a given integration time. The TIA feedback resistance determines gain, bandwidth, and noise, with bandwidth and noise decreasing with larger resistance\cite{razavi_sscm2019}. Compact waveguide-coupled photodiodes and tight integration between the photodiodes and TIAs yields a small parasitic capacitance of $1.5~\mathrm{fF}$. We thus achieve low noise in the electrical signal chain for $20~\mathrm{k\Omega}$ gain and bandwidths above $280~\mathrm{MHz}$, as shown in Fig. \ref{fig:3_electrooptic_perf}(a)-(b). As seen in Fig. \ref{fig:3_electrooptic_perf}(d), our integrated TIA has a similar gain-bandwidth product and a $2-3 \times$ lower noise floor than conventional systems with photodiodes and amplifiers on separate chips.

The low noise on-chip amplifier chain allows the receiver FPA to operate at the quantum limit for sensitivity, which is reached when local oscillator shot noise dominates all other noise sources\cite{mjcollett_jmo1987, marubin_ol2007}. As shown in Fig. \ref{fig:3_electrooptic_perf}(e), shot noise reaches parity with amplifier noise with only $5~\mathrm{\mu W}$ of LO power for a typical receiver pixel. In contrast, coherent receivers for telecommunications typically require $\sim 10-100 \times$ more LO power. Combined with the excellent $30-40~\mathrm{dB}$ common-mode rejection ratio of the balanced heterodyne detectors, as shown in Fig. \ref{fig:3_electrooptic_perf}(c), this makes the receiver array significantly less susceptible to LO noise sources such as laser relative intensity noise, optical amplifier noise, and chirp generator noise. Furthermore, the low LO power per pixel reduces the size of the LO distribution tree since many receiver pixels can simultaneously share LO power.

Monolithic integration of electronics into the receiver FPA facilitates the use of an actively multiplexed readout architecture, allowing the receiver to be scaled to arbitrarily large numbers of pixels. In our demonstrator chip, multiple levels of multiplexing and amplification are used to map 512 pixels to 8 outputs while maintaining signal integrity. As illustrated in Extended Data Fig. \ref{extfig:2_readout_arch_and_synch}(a-c), the pixels are read out in blocks of 8 at a time. The output analog signals are fed into a bank of off-chip analog-to-digital converters for digitization, followed by digital signal processing on a field-programmable gate-array (FPGA).

The transmitter illumination pattern is closely synchronized with the receiver readout pattern, as detailed in the Methods and Extended Data Fig. \ref{extfig:2_readout_arch_and_synch}(d). To further improve the optical efficiency of the system\cite{jywang_ao1982}, a microlens array was placed in front of the transmitter array as illustrated in Extended Data Fig. \ref{extfig:3_optical_schematic}. This produced a structured illumination pattern that exactly matched the grating coupler positions in the receiver FPA, as shown in Extended Data Fig. \ref{extfig:4_farfield}, yielding a $24\times$ improvement in signal strength.

\section*{3D imaging and velocimetry}
\begin{figure*}[p]
\centering
\includegraphics[width=\textwidth]{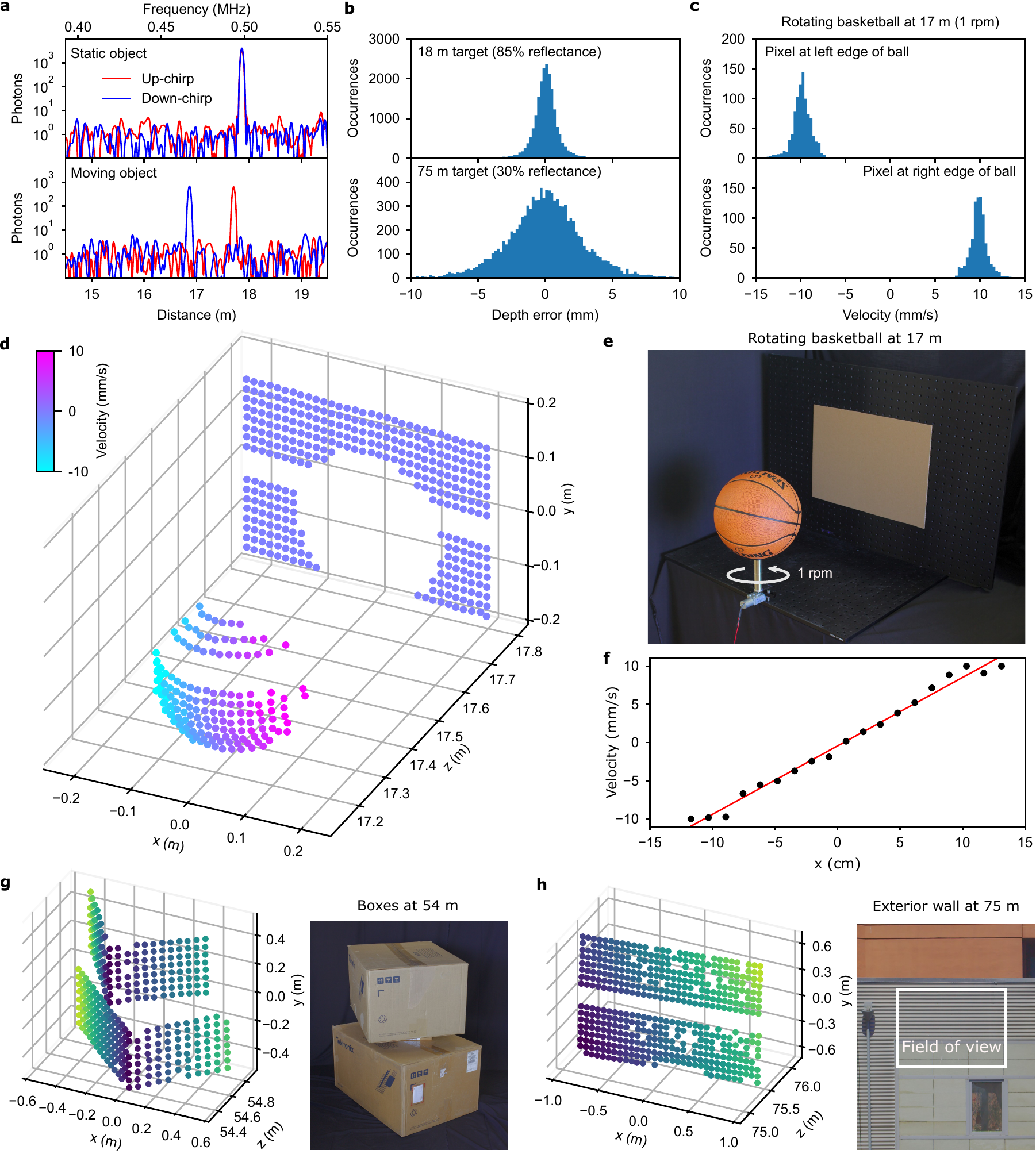}
\caption{3D imaging system characterization. (a) Representative signals from a receiver pixel, showing Doppler splitting between the up- and down-chirps for the moving target. (b) Depth noise for targets at $18~\mathrm{m}$ and $75~\mathrm{m}$, with standard deviations of $1.8~\mathrm{mm}$ and $3.1~\mathrm{mm}$ respectively. (c) Velocity histograms for a basketball rotating at $1~\mathrm{rpm}$, exhibiting a standard deviation of $1.0~\mathrm{mm}/\mathrm{s}$. (d) Velocity annotated point cloud of a basketball at $17~\mathrm{m}$ rotating about its vertical axis at $1~\mathrm{rpm}$. (e) Photograph of the basketball setup. (f) Horizontal linecut of velocity across the middle of the basketball. (g, h) Point clouds of (g) stacked cardboard boxes at $54~\mathrm{m}$, and (h) an exterior wall at $75~\mathrm{m}$. Distance to the target is indicated by colour in (g) and (h). The missing band of points in the middle of the point clouds is due to a narrow gap in the receiver array for electrical and optical routing.}
\label{fig:4_results}
\end{figure*}

LiDAR system operation is presented in Fig. \ref{fig:4_results} using an emitter power of $4~\mathrm{mW}$ at the aperture, a chirp bandwidth of $4~\mathrm{GHz}$, and up- and down-chirp lengths of $850~\mathrm{\mu s}$. As shown in Fig. \ref{fig:4_results}(a), distance and velocity are encoded in the frequency tones detected by each pixel\cite{hdgriffiths_ecej1990, jriemensberger_nature2020}. As demonstrated in Fig. \ref{fig:4_results}(b), the system achieved a measurement precision of $1.8~\mathrm{mm}$ at $17~\mathrm{m}$ for a $85\%$ reflectance target, and $3.1~\mathrm{mm}$ at $75~\mathrm{m}$ for a $30\%$ reflectance target. Due to the effects of speckle, which equally impacts all coherent radar and LiDAR schemes\cite{jywang_ao1982}, the detection probability was $97\%$ for the $17~\mathrm{m}$ target, and $42\%$ for the $75~\mathrm{m}$ target. Meanwhile, the velocity precision for slowly moving objects is $1.0~\mathrm{mm}/\mathrm{s}$, as shown in Fig. \ref{fig:4_results}(c). Point clouds of a rotating basketball at $17~\mathrm{m}$, stacked boxes at $55~\mathrm{m}$, and an exterior wall at $75~\mathrm{m}$ are illustrated in Fig. \ref{fig:4_results}(d-h). The point clouds were generated by stacking 3 sequential frames to minimize the number of missing pixels due to speckle effects. No incoherent averaging was used. Detection probabilities were calculated without any frame stacking. The missing band of points in the middle of the point clouds is due to a narrow gap in the receiver array for electrical and optical routing, as shown in Fig. \ref{fig:1_architecture}(c). In future designs, this gap can be reduced or eliminated by more aggressive chip layout and routing.

\section*{Discussion and outlook}
We have demonstrated a scalable solid-state 3D imaging architecture that achieves $>70~\mathrm{m}$ range and millimetre-class accuracy with only $4~\mathrm{mW}$ of transmitted power. Our $3.1~\mathrm{mm}$ precision is an order of magnitude higher than existing solid-state 3D imagers at these ranges, with state-of-the-art flash LiDAR systems limited to an accuracy of several centimetres for distances greater than 50 metres\cite{pmcmanamon_oe2012, pfmcmanamon_oe2017, swhutchings_ijssc2019, arximenes_ijscc2019}. This level of performance meets the needs of a variety of demanding applications previously out of reach for solid state 3D imaging systems. The long range and eye safety requirements of autonomous driving\cite{curmson_jofr2008}, and the millimeter-class accuracy required by 3D mapping of constructions sites and buildings\cite{drebolj_ac2017, jwang_jirs2015, akasturi_lrta2016, qwang_aei2019} can be simultaneously achieved by our system.

System range could be improved by correcting the $4\times$ mismatch between the number of transmitter and receiver positions as discussed in Methods, and through increases in transmitter power. The latter can be achieved by optimizing the silicon photonic elements to minimize transmission losses. Reducing two-photon absorption with wider waveguides or reverse-biased PIN junctions would further increase transmitter power, with previous demonstrations reaching $\sim 1~\mathrm{W}$ optical power \cite{hrong_nature2005}. Since the range of a coherent LiDAR scales as the square root of transmitter power\cite{jywang_ao1982}, this implies that our architecture could operate at ranges of up to $1~\mathrm{km}$. Conversely, for a fixed distance of $75~\mathrm{m}$, increasing the transmitter power to $1~\mathrm{W}$ and eliminating the mismatch between transmitter and receiver would increase the point rate from $4.7 \times 10^3$ to $5 \times 10^6$ points per second. Finally, depth accuracy could be improved by increasing chirp bandwidth. Demonstrations of $50~\mathrm{GHz}$ silicon photonic modulators \cite{asamani_jlwt2019} imply that $\sim200~\mathrm{\mu m}$ precision is feasible.

Our 3D imaging architecture naturally scales to large arrays. By using a $1 \times N$ switching tree to steer light, the number of active thermo-optic switches and associated DACs and drivers needed for control is $O(\log N)$. On the receiver side, power consumption depends only upon the number of parallel readout channels and is independent of the size of the array since only active amplifiers are powered at any given time. The fundamental limit on array size therefore comes from the size of the chip. At the current receiver pixel pitch of $80 \times 100 ~\mathrm{\mu m}^2$, chips the size of a full-frame camera sensor ($36 \times 24 ~\mathrm{mm}^2$) would have QVGA ($320 \times 240$ pixel) resolution. However, using state-of-the-art designs $8 \times 5~\mathrm{\mu m}^2$ pixels are feasible, corresponding to full-frame sensors with resolutions of $4500 \times 4800$ pixels. Further design refinements could yield even higher resolutions.

We have developed a universal solid-state 3D imaging architecture with the potential to meet the needs of nearly all 3D imaging applications, spanning from robotics and autonomous navigation to consumer products such as augmented reality headsets. Our results suggest that the equivalent of the CMOS image sensor for 3D imaging is imminent, ushering in a broad range of applications which were previously impractical or unimaginable.

\subsection*{Acknowledgements}
D. J. Thomson acknowledges funding from the Royal Society for his University Research Fellowship. We thank Andy Stricker, Andy Watts, Mehrdad Djavid and the rest of the Global Foundries team for assistance in device fabrication. 

\subsection*{Contributions}
C.R. and A.Y.P. contributed equally to this work. C.R. conceived, built and tested the free-space portion of the LiDAR system, performed the final LiDAR measurements, and calibrated the optical switching trees. A.Y.P and C.R. performed the electro-optic characterization. A.Y.P. conceived, built and tested the fiber-optic portion of the LiDAR system, wrote the LiDAR system software, tested the coherent receiver array, and contributed to the architecture and layout of the photonic chip. D.J.T. designed and laid out the silicon photonics modulator. R.W. contributed to design verification including circuit simulations, and contributed to the embedded software control systems. I.O. designed the electronic circuits on the photonic chip and performed their layout and verification. S.A.F developed the signal acquisition and control systems, and contributed to the system architecture. A.J.C. designed and verified the circuit boards used to interface with the photonic chip. A.G. contributed to the architecture and performed the layout of the photonic chip. F.M and X.C. contributed to the fabrication and testing of the modulator. R.N. conceived the receiver, switching, and photonic system architecture. R.N. and G.T.R. supervised the project.
 
\subsection*{Competing Interests} In terms of competing interests, all authors with the exception of F.M. and X.C. disclose that they are shareholders of Pointcloud Inc., a start-up company engaged in making laser ranging devices based on coherent receiver arrays.

%% Put the bibliography here, most people will use BiBTeX in
%% which case the environment below should be replaced with
%% the \bibliography{} command.
\bibliography{lidar_refs.bib}

\begin{thebibliography}{50}%
\makeatletter
\providecommand \@ifxundefined [1]{%
 \@ifx{#1\undefined}
}%
\providecommand \@ifnum [1]{%
 \ifnum #1\expandafter \@firstoftwo
 \else \expandafter \@secondoftwo
 \fi
}%
\providecommand \@ifx [1]{%
 \ifx #1\expandafter \@firstoftwo
 \else \expandafter \@secondoftwo
 \fi
}%
\providecommand \natexlab [1]{#1}%
\providecommand \enquote  [1]{``#1''}%
\providecommand \bibnamefont  [1]{#1}%
\providecommand \bibfnamefont [1]{#1}%
\providecommand \citenamefont [1]{#1}%
\providecommand \href@noop [0]{\@secondoftwo}%
\providecommand \href [0]{\begingroup \@sanitize@url \@href}%
\providecommand \@href[1]{\@@startlink{#1}\@@href}%
\providecommand \@@href[1]{\endgroup#1\@@endlink}%
\providecommand \@sanitize@url [0]{\catcode `\\12\catcode `\$12\catcode
  `\&12\catcode `\#12\catcode `\^12\catcode `\_12\catcode `\%12\relax}%
\providecommand \@@startlink[1]{}%
\providecommand \@@endlink[0]{}%
\providecommand \url  [0]{\begingroup\@sanitize@url \@url }%
\providecommand \@url [1]{\endgroup\@href {#1}{\urlprefix }}%
\providecommand \urlprefix  [0]{URL }%
\providecommand \Eprint [0]{\href }%
\providecommand \doibase [0]{http://dx.doi.org/}%
\providecommand \selectlanguage [0]{\@gobble}%
\providecommand \bibinfo  [0]{\@secondoftwo}%
\providecommand \bibfield  [0]{\@secondoftwo}%
\providecommand \translation [1]{[#1]}%
\providecommand \BibitemOpen [0]{}%
\providecommand \bibitemStop [0]{}%
\providecommand \bibitemNoStop [0]{.\EOS\space}%
\providecommand \EOS [0]{\spacefactor3000\relax}%
\providecommand \BibitemShut  [1]{\csname bibitem#1\endcsname}%
\let\auto@bib@innerbib\@empty
%</preamble>
\bibitem [{\citenamefont {Urmson}\ \emph {et~al.}(2008)\citenamefont {Urmson},
  \citenamefont {Anhalt}, \citenamefont {Bagnell}, \citenamefont {Baker},
  \citenamefont {Bittner}, \citenamefont {Clark}, \citenamefont {Dolan},
  \citenamefont {Duggins}, \citenamefont {Galatali}, \citenamefont {Geyer},
  \citenamefont {Gittleman}, \citenamefont {Harbaugh}, \citenamefont {Hebert},
  \citenamefont {Howard}, \citenamefont {Kolski}, \citenamefont {Kelly},
  \citenamefont {Likhachev}, \citenamefont {McNaughton}, \citenamefont
  {Miller}, \citenamefont {Peterson}, \citenamefont {Pilnick}, \citenamefont
  {Rajkumar}, \citenamefont {Rybski}, \citenamefont {Salesky}, \citenamefont
  {Seo}, \citenamefont {Singh}, \citenamefont {Snider}, \citenamefont {Stentz},
  \citenamefont {Whittaker}, \citenamefont {Wolkowicki}, \citenamefont
  {Ziglar}, \citenamefont {Bae}, \citenamefont {Brown}, \citenamefont
  {Demitrish}, \citenamefont {Litkouhi}, \citenamefont {Nickolaou},
  \citenamefont {Sadekar}, \citenamefont {Zhang}, \citenamefont {Struble},
  \citenamefont {Taylor}, \citenamefont {Darms},\ and\ \citenamefont
  {Ferguson}}]{curmson_jofr2008}%
  \BibitemOpen
  \bibfield  {author} {\bibinfo {author} {\bibfnamefont {C.}~\bibnamefont
  {Urmson}}, \bibinfo {author} {\bibfnamefont {J.}~\bibnamefont {Anhalt}},
  \bibinfo {author} {\bibfnamefont {D.}~\bibnamefont {Bagnell}}, \bibinfo
  {author} {\bibfnamefont {C.}~\bibnamefont {Baker}}, \bibinfo {author}
  {\bibfnamefont {R.}~\bibnamefont {Bittner}}, \bibinfo {author} {\bibfnamefont
  {M.~N.}\ \bibnamefont {Clark}}, \bibinfo {author} {\bibfnamefont
  {J.}~\bibnamefont {Dolan}}, \bibinfo {author} {\bibfnamefont
  {D.}~\bibnamefont {Duggins}}, \bibinfo {author} {\bibfnamefont
  {T.}~\bibnamefont {Galatali}}, \bibinfo {author} {\bibfnamefont
  {C.}~\bibnamefont {Geyer}}, \bibinfo {author} {\bibfnamefont
  {M.}~\bibnamefont {Gittleman}}, \bibinfo {author} {\bibfnamefont
  {S.}~\bibnamefont {Harbaugh}}, \bibinfo {author} {\bibfnamefont
  {M.}~\bibnamefont {Hebert}}, \bibinfo {author} {\bibfnamefont {T.~M.}\
  \bibnamefont {Howard}}, \bibinfo {author} {\bibfnamefont {S.}~\bibnamefont
  {Kolski}}, \bibinfo {author} {\bibfnamefont {A.}~\bibnamefont {Kelly}},
  \bibinfo {author} {\bibfnamefont {M.}~\bibnamefont {Likhachev}}, \bibinfo
  {author} {\bibfnamefont {M.}~\bibnamefont {McNaughton}}, \bibinfo {author}
  {\bibfnamefont {N.}~\bibnamefont {Miller}}, \bibinfo {author} {\bibfnamefont
  {K.}~\bibnamefont {Peterson}}, \bibinfo {author} {\bibfnamefont
  {B.}~\bibnamefont {Pilnick}}, \bibinfo {author} {\bibfnamefont
  {R.}~\bibnamefont {Rajkumar}}, \bibinfo {author} {\bibfnamefont
  {P.}~\bibnamefont {Rybski}}, \bibinfo {author} {\bibfnamefont
  {B.}~\bibnamefont {Salesky}}, \bibinfo {author} {\bibfnamefont {Y.-W.}\
  \bibnamefont {Seo}}, \bibinfo {author} {\bibfnamefont {S.}~\bibnamefont
  {Singh}}, \bibinfo {author} {\bibfnamefont {J.}~\bibnamefont {Snider}},
  \bibinfo {author} {\bibfnamefont {A.}~\bibnamefont {Stentz}}, \bibinfo
  {author} {\bibfnamefont {W.~R.}\ \bibnamefont {Whittaker}}, \bibinfo {author}
  {\bibfnamefont {Z.}~\bibnamefont {Wolkowicki}}, \bibinfo {author}
  {\bibfnamefont {J.}~\bibnamefont {Ziglar}}, \bibinfo {author} {\bibfnamefont
  {H.}~\bibnamefont {Bae}}, \bibinfo {author} {\bibfnamefont {T.}~\bibnamefont
  {Brown}}, \bibinfo {author} {\bibfnamefont {D.}~\bibnamefont {Demitrish}},
  \bibinfo {author} {\bibfnamefont {B.}~\bibnamefont {Litkouhi}}, \bibinfo
  {author} {\bibfnamefont {J.}~\bibnamefont {Nickolaou}}, \bibinfo {author}
  {\bibfnamefont {V.}~\bibnamefont {Sadekar}}, \bibinfo {author} {\bibfnamefont
  {W.}~\bibnamefont {Zhang}}, \bibinfo {author} {\bibfnamefont
  {J.}~\bibnamefont {Struble}}, \bibinfo {author} {\bibfnamefont
  {M.}~\bibnamefont {Taylor}}, \bibinfo {author} {\bibfnamefont
  {M.}~\bibnamefont {Darms}}, \ and\ \bibinfo {author} {\bibfnamefont
  {D.}~\bibnamefont {Ferguson}},\ }\href {\doibase 10.1002/rob.20255}
  {\bibfield  {journal} {\bibinfo  {journal} {Journal of Field Robotics}\
  }\textbf {\bibinfo {volume} {25}},\ \bibinfo {pages} {425} (\bibinfo {year}
  {2008})}\BibitemShut {NoStop}%
\bibitem [{\citenamefont {Wang}\ and\ \citenamefont
  {Kim}(2019)}]{qwang_aei2019}%
  \BibitemOpen
  \bibfield  {author} {\bibinfo {author} {\bibfnamefont {Q.}~\bibnamefont
  {Wang}}\ and\ \bibinfo {author} {\bibfnamefont {M.-K.}\ \bibnamefont {Kim}},\
  }\href {\doibase 10.1016/j.aei.2019.02.007} {\bibfield  {journal} {\bibinfo
  {journal} {Advanced Engineering Informatics}\ }\textbf {\bibinfo {volume}
  {39}},\ \bibinfo {pages} {306 } (\bibinfo {year} {2019})}\BibitemShut
  {NoStop}%
\bibitem [{\citenamefont {Lichti}(2007)}]{ddlichti_ijprs2007}%
  \BibitemOpen
  \bibfield  {author} {\bibinfo {author} {\bibfnamefont {D.~D.}\ \bibnamefont
  {Lichti}},\ }\href {\doibase 10.1016/j.isprsjprs.2006.10.004} {\bibfield
  {journal} {\bibinfo  {journal} {ISPRS Journal of Photogrammetry and Remote
  Sensing}\ }\textbf {\bibinfo {volume} {61}},\ \bibinfo {pages} {307}
  (\bibinfo {year} {2007})}\BibitemShut {NoStop}%
\bibitem [{\citenamefont {Kidd}(2017)}]{jkidd_mscthesis2017}%
  \BibitemOpen
  \bibfield  {author} {\bibinfo {author} {\bibfnamefont {J.}~\bibnamefont
  {Kidd}},\ }\emph {\bibinfo {title} {Performance Evaluation of the Velodyne
  VLP-16 System for Surface Feature Surveying}},\ \href
  {https://scholars.unh.edu/thesis/1116} {Master's thesis},\ \bibinfo  {school}
  {University of New Hampshire} (\bibinfo {year} {2017})\BibitemShut {NoStop}%
\bibitem [{\citenamefont {Salvi}\ \emph {et~al.}(2004)\citenamefont {Salvi},
  \citenamefont {Pagès},\ and\ \citenamefont {Batlle}}]{jsalvi_pr2004}%
  \BibitemOpen
  \bibfield  {author} {\bibinfo {author} {\bibfnamefont {J.}~\bibnamefont
  {Salvi}}, \bibinfo {author} {\bibfnamefont {J.}~\bibnamefont {Pagès}}, \
  and\ \bibinfo {author} {\bibfnamefont {J.}~\bibnamefont {Batlle}},\ }\href
  {\doibase 10.1016/j.patcog.2003.10.002} {\bibfield  {journal} {\bibinfo
  {journal} {Pattern Recognition}\ }\textbf {\bibinfo {volume} {37}},\ \bibinfo
  {pages} {827 } (\bibinfo {year} {2004})},\ \bibinfo {note} {agent Based
  Computer Vision}\BibitemShut {NoStop}%
\bibitem [{\citenamefont {Corti}\ \emph {et~al.}(2016)\citenamefont {Corti},
  \citenamefont {Giancola}, \citenamefont {Mainetti},\ and\ \citenamefont
  {Sala}}]{acorti_ras2016}%
  \BibitemOpen
  \bibfield  {author} {\bibinfo {author} {\bibfnamefont {A.}~\bibnamefont
  {Corti}}, \bibinfo {author} {\bibfnamefont {S.}~\bibnamefont {Giancola}},
  \bibinfo {author} {\bibfnamefont {G.}~\bibnamefont {Mainetti}}, \ and\
  \bibinfo {author} {\bibfnamefont {R.}~\bibnamefont {Sala}},\ }\href {\doibase
  10.1016/j.robot.2015.09.024} {\bibfield  {journal} {\bibinfo  {journal}
  {Robotics and Autonomous Systems}\ }\textbf {\bibinfo {volume} {75}},\
  \bibinfo {pages} {584} (\bibinfo {year} {2016})}\BibitemShut {NoStop}%
\bibitem [{\citenamefont {McManamon}(2012)}]{pmcmanamon_oe2012}%
  \BibitemOpen
  \bibfield  {author} {\bibinfo {author} {\bibfnamefont {P.}~\bibnamefont
  {McManamon}},\ }\href {\doibase 10.1117/1.OE.51.6.060901} {\bibfield
  {journal} {\bibinfo  {journal} {Optical Engineering}\ }\textbf {\bibinfo
  {volume} {51}},\ \bibinfo {pages} {1 } (\bibinfo {year} {2012})}\BibitemShut
  {NoStop}%
\bibitem [{\citenamefont {McManamon}\ \emph {et~al.}(2017)\citenamefont
  {McManamon}, \citenamefont {Banks}, \citenamefont {Beck}, \citenamefont
  {Fried}, \citenamefont {Huntington},\ and\ \citenamefont
  {Watson}}]{pfmcmanamon_oe2017}%
  \BibitemOpen
  \bibfield  {author} {\bibinfo {author} {\bibfnamefont {P.~F.}\ \bibnamefont
  {McManamon}}, \bibinfo {author} {\bibfnamefont {P.~S.}\ \bibnamefont
  {Banks}}, \bibinfo {author} {\bibfnamefont {J.~D.}\ \bibnamefont {Beck}},
  \bibinfo {author} {\bibfnamefont {D.~G.}\ \bibnamefont {Fried}}, \bibinfo
  {author} {\bibfnamefont {A.~S.}\ \bibnamefont {Huntington}}, \ and\ \bibinfo
  {author} {\bibfnamefont {E.~A.}\ \bibnamefont {Watson}},\ }\href {\doibase
  10.1117/1.OE.56.3.031223} {\bibfield  {journal} {\bibinfo  {journal} {Optical
  Engineering}\ }\textbf {\bibinfo {volume} {56}},\ \bibinfo {pages} {1 }
  (\bibinfo {year} {2017})}\BibitemShut {NoStop}%
\bibitem [{\citenamefont {{Hutchings}}\ \emph {et~al.}(2019)\citenamefont
  {{Hutchings}}, \citenamefont {{Johnston}}, \citenamefont {{Gyongy}},
  \citenamefont {{Al Abbas}}, \citenamefont {{Dutton}}, \citenamefont
  {{Tyler}}, \citenamefont {{Chan}}, \citenamefont {{Leach}},\ and\
  \citenamefont {{Henderson}}}]{swhutchings_ijssc2019}%
  \BibitemOpen
  \bibfield  {author} {\bibinfo {author} {\bibfnamefont {S.~W.}\ \bibnamefont
  {{Hutchings}}}, \bibinfo {author} {\bibfnamefont {N.}~\bibnamefont
  {{Johnston}}}, \bibinfo {author} {\bibfnamefont {I.}~\bibnamefont
  {{Gyongy}}}, \bibinfo {author} {\bibfnamefont {T.}~\bibnamefont {{Al
  Abbas}}}, \bibinfo {author} {\bibfnamefont {N.~A.~W.}\ \bibnamefont
  {{Dutton}}}, \bibinfo {author} {\bibfnamefont {M.}~\bibnamefont {{Tyler}}},
  \bibinfo {author} {\bibfnamefont {S.}~\bibnamefont {{Chan}}}, \bibinfo
  {author} {\bibfnamefont {J.}~\bibnamefont {{Leach}}}, \ and\ \bibinfo
  {author} {\bibfnamefont {R.~K.}\ \bibnamefont {{Henderson}}},\ }\href
  {\doibase 10.1109/JSSC.2019.2939083} {\bibfield  {journal} {\bibinfo
  {journal} {IEEE Journal of Solid-State Circuits}\ }\textbf {\bibinfo {volume}
  {54}},\ \bibinfo {pages} {2947} (\bibinfo {year} {2019})}\BibitemShut
  {NoStop}%
\bibitem [{\citenamefont {{Ronchini Ximenes}}\ \emph
  {et~al.}(2019)\citenamefont {{Ronchini Ximenes}}, \citenamefont
  {{Padmanabhan}}, \citenamefont {{Lee}}, \citenamefont {{Yamashita}},
  \citenamefont {{Yaung}},\ and\ \citenamefont
  {{Charbon}}}]{arximenes_ijscc2019}%
  \BibitemOpen
  \bibfield  {author} {\bibinfo {author} {\bibfnamefont {A.}~\bibnamefont
  {{Ronchini Ximenes}}}, \bibinfo {author} {\bibfnamefont {P.}~\bibnamefont
  {{Padmanabhan}}}, \bibinfo {author} {\bibfnamefont {M.}~\bibnamefont
  {{Lee}}}, \bibinfo {author} {\bibfnamefont {Y.}~\bibnamefont {{Yamashita}}},
  \bibinfo {author} {\bibfnamefont {D.}~\bibnamefont {{Yaung}}}, \ and\
  \bibinfo {author} {\bibfnamefont {E.}~\bibnamefont {{Charbon}}},\ }\href
  {\doibase 10.1109/JSSC.2019.2938412} {\bibfield  {journal} {\bibinfo
  {journal} {IEEE Journal of Solid-State Circuits}\ }\textbf {\bibinfo {volume}
  {54}},\ \bibinfo {pages} {3203} (\bibinfo {year} {2019})}\BibitemShut
  {NoStop}%
\bibitem [{\citenamefont {{Behroozpour}}\ \emph {et~al.}(2017)\citenamefont
  {{Behroozpour}}, \citenamefont {{Sandborn}}, \citenamefont {{Wu}},\ and\
  \citenamefont {{Boser}}}]{bbehroozpour_ieeecm2017}%
  \BibitemOpen
  \bibfield  {author} {\bibinfo {author} {\bibfnamefont {B.}~\bibnamefont
  {{Behroozpour}}}, \bibinfo {author} {\bibfnamefont {P.~A.~M.}\ \bibnamefont
  {{Sandborn}}}, \bibinfo {author} {\bibfnamefont {M.~C.}\ \bibnamefont
  {{Wu}}}, \ and\ \bibinfo {author} {\bibfnamefont {B.~E.}\ \bibnamefont
  {{Boser}}},\ }\href {\doibase 10.1109/MCOM.2017.1700030} {\bibfield
  {journal} {\bibinfo  {journal} {IEEE Communications Magazine}\ }\textbf
  {\bibinfo {volume} {55}},\ \bibinfo {pages} {135} (\bibinfo {year}
  {2017})}\BibitemShut {NoStop}%
\bibitem [{\citenamefont {Aflatouni}\ \emph {et~al.}(2015)\citenamefont
  {Aflatouni}, \citenamefont {Abiri}, \citenamefont {Rekhi},\ and\
  \citenamefont {Hajimiri}}]{faflatouni_oe2015}%
  \BibitemOpen
  \bibfield  {author} {\bibinfo {author} {\bibfnamefont {F.}~\bibnamefont
  {Aflatouni}}, \bibinfo {author} {\bibfnamefont {B.}~\bibnamefont {Abiri}},
  \bibinfo {author} {\bibfnamefont {A.}~\bibnamefont {Rekhi}}, \ and\ \bibinfo
  {author} {\bibfnamefont {A.}~\bibnamefont {Hajimiri}},\ }\href {\doibase
  10.1364/OE.23.005117} {\bibfield  {journal} {\bibinfo  {journal} {Opt.
  Express}\ }\textbf {\bibinfo {volume} {23}},\ \bibinfo {pages} {5117}
  (\bibinfo {year} {2015})}\BibitemShut {NoStop}%
\bibitem [{\citenamefont {{Martin}}\ \emph {et~al.}(2018)\citenamefont
  {{Martin}}, \citenamefont {{Dodane}}, \citenamefont {{Leviandier}},
  \citenamefont {{Dolfi}}, \citenamefont {{Naughton}}, \citenamefont
  {{O’Brien}}, \citenamefont {{Spuessens}}, \citenamefont {{Baets}},
  \citenamefont {{Lepage}}, \citenamefont {{Verheyen}}, \citenamefont {{De
  Heyn}}, \citenamefont {{Absil}}, \citenamefont {{Feneyrou}},\ and\
  \citenamefont {{Bourderionnet}}}]{amartin_jlt2018}%
  \BibitemOpen
  \bibfield  {author} {\bibinfo {author} {\bibfnamefont {A.}~\bibnamefont
  {{Martin}}}, \bibinfo {author} {\bibfnamefont {D.}~\bibnamefont {{Dodane}}},
  \bibinfo {author} {\bibfnamefont {L.}~\bibnamefont {{Leviandier}}}, \bibinfo
  {author} {\bibfnamefont {D.}~\bibnamefont {{Dolfi}}}, \bibinfo {author}
  {\bibfnamefont {A.}~\bibnamefont {{Naughton}}}, \bibinfo {author}
  {\bibfnamefont {P.}~\bibnamefont {{O’Brien}}}, \bibinfo {author}
  {\bibfnamefont {T.}~\bibnamefont {{Spuessens}}}, \bibinfo {author}
  {\bibfnamefont {R.}~\bibnamefont {{Baets}}}, \bibinfo {author} {\bibfnamefont
  {G.}~\bibnamefont {{Lepage}}}, \bibinfo {author} {\bibfnamefont
  {P.}~\bibnamefont {{Verheyen}}}, \bibinfo {author} {\bibfnamefont
  {P.}~\bibnamefont {{De Heyn}}}, \bibinfo {author} {\bibfnamefont
  {P.}~\bibnamefont {{Absil}}}, \bibinfo {author} {\bibfnamefont
  {P.}~\bibnamefont {{Feneyrou}}}, \ and\ \bibinfo {author} {\bibfnamefont
  {J.}~\bibnamefont {{Bourderionnet}}},\ }\href {\doibase
  10.1109/JLT.2018.2840223} {\bibfield  {journal} {\bibinfo  {journal} {Journal
  of Lightwave Technology}\ }\textbf {\bibinfo {volume} {36}},\ \bibinfo
  {pages} {4640} (\bibinfo {year} {2018})}\BibitemShut {NoStop}%
\bibitem [{\citenamefont {Inoue}\ \emph {et~al.}(2019)\citenamefont {Inoue},
  \citenamefont {Ichikawa}, \citenamefont {Kawasaki},\ and\ \citenamefont
  {Yamashita}}]{dinoue_oe2019}%
  \BibitemOpen
  \bibfield  {author} {\bibinfo {author} {\bibfnamefont {D.}~\bibnamefont
  {Inoue}}, \bibinfo {author} {\bibfnamefont {T.}~\bibnamefont {Ichikawa}},
  \bibinfo {author} {\bibfnamefont {A.}~\bibnamefont {Kawasaki}}, \ and\
  \bibinfo {author} {\bibfnamefont {T.}~\bibnamefont {Yamashita}},\ }\href
  {\doibase 10.1364/OE.27.002499} {\bibfield  {journal} {\bibinfo  {journal}
  {Opt. Express}\ }\textbf {\bibinfo {volume} {27}},\ \bibinfo {pages} {2499}
  (\bibinfo {year} {2019})}\BibitemShut {NoStop}%
\bibitem [{\citenamefont {Li}\ \emph {et~al.}(2019)\citenamefont {Li},
  \citenamefont {Cao}, \citenamefont {Wu}, \citenamefont {Li},\ and\
  \citenamefont {Chen}}]{cli_oe2019}%
  \BibitemOpen
  \bibfield  {author} {\bibinfo {author} {\bibfnamefont {C.}~\bibnamefont
  {Li}}, \bibinfo {author} {\bibfnamefont {X.}~\bibnamefont {Cao}}, \bibinfo
  {author} {\bibfnamefont {K.}~\bibnamefont {Wu}}, \bibinfo {author}
  {\bibfnamefont {X.}~\bibnamefont {Li}}, \ and\ \bibinfo {author}
  {\bibfnamefont {J.}~\bibnamefont {Chen}},\ }\href {\doibase
  10.1364/OE.27.032970} {\bibfield  {journal} {\bibinfo  {journal} {Opt.
  Express}\ }\textbf {\bibinfo {volume} {27}},\ \bibinfo {pages} {32970}
  (\bibinfo {year} {2019})}\BibitemShut {NoStop}%
\bibitem [{\citenamefont {Collett}\ \emph {et~al.}(1987)\citenamefont
  {Collett}, \citenamefont {Loudon},\ and\ \citenamefont
  {Gardiner}}]{mjcollett_jmo1987}%
  \BibitemOpen
  \bibfield  {author} {\bibinfo {author} {\bibfnamefont {M.}~\bibnamefont
  {Collett}}, \bibinfo {author} {\bibfnamefont {R.}~\bibnamefont {Loudon}}, \
  and\ \bibinfo {author} {\bibfnamefont {C.}~\bibnamefont {Gardiner}},\ }\href
  {\doibase 10.1080/09500348714550811} {\bibfield  {journal} {\bibinfo
  {journal} {Journal of Modern Optics}\ }\textbf {\bibinfo {volume} {34}},\
  \bibinfo {pages} {881} (\bibinfo {year} {1987})}\BibitemShut {NoStop}%
\bibitem [{\citenamefont {Rubin}\ and\ \citenamefont
  {Kaushik}(2007)}]{marubin_ol2007}%
  \BibitemOpen
  \bibfield  {author} {\bibinfo {author} {\bibfnamefont {M.~A.}\ \bibnamefont
  {Rubin}}\ and\ \bibinfo {author} {\bibfnamefont {S.}~\bibnamefont
  {Kaushik}},\ }\href {\doibase 10.1364/OL.32.001369} {\bibfield  {journal}
  {\bibinfo  {journal} {Opt. Lett.}\ }\textbf {\bibinfo {volume} {32}},\
  \bibinfo {pages} {1369} (\bibinfo {year} {2007})}\BibitemShut {NoStop}%
\bibitem [{\citenamefont {{El Gamal}}\ and\ \citenamefont
  {{Eltoukhy}}(2005)}]{aelgamal_ieeecdm2005}%
  \BibitemOpen
  \bibfield  {author} {\bibinfo {author} {\bibfnamefont {A.}~\bibnamefont {{El
  Gamal}}}\ and\ \bibinfo {author} {\bibfnamefont {H.}~\bibnamefont
  {{Eltoukhy}}},\ }\href {\doibase 10.1109/MCD.2005.1438751} {\bibfield
  {journal} {\bibinfo  {journal} {IEEE Circuits and Devices Magazine}\ }\textbf
  {\bibinfo {volume} {21}},\ \bibinfo {pages} {6} (\bibinfo {year}
  {2005})}\BibitemShut {NoStop}%
\bibitem [{\citenamefont {Stann}\ \emph {et~al.}(2004)\citenamefont {Stann},
  \citenamefont {Aliberti}, \citenamefont {Carothers}, \citenamefont {Dammann},
  \citenamefont {Dang}, \citenamefont {Giza}, \citenamefont {Lawler},
  \citenamefont {Redman},\ and\ \citenamefont {Simon}}]{blstann_lrta2004}%
  \BibitemOpen
  \bibfield  {author} {\bibinfo {author} {\bibfnamefont {B.~L.}\ \bibnamefont
  {Stann}}, \bibinfo {author} {\bibfnamefont {K.}~\bibnamefont {Aliberti}},
  \bibinfo {author} {\bibfnamefont {D.}~\bibnamefont {Carothers}}, \bibinfo
  {author} {\bibfnamefont {J.}~\bibnamefont {Dammann}}, \bibinfo {author}
  {\bibfnamefont {G.}~\bibnamefont {Dang}}, \bibinfo {author} {\bibfnamefont
  {M.~M.}\ \bibnamefont {Giza}}, \bibinfo {author} {\bibfnamefont {W.~B.}\
  \bibnamefont {Lawler}}, \bibinfo {author} {\bibfnamefont {B.~C.}\
  \bibnamefont {Redman}}, \ and\ \bibinfo {author} {\bibfnamefont {D.~R.}\
  \bibnamefont {Simon}},\ }in\ \href {\doibase 10.1117/12.542549} {\emph
  {\bibinfo {booktitle} {Laser Radar Technology and Applications IX}}},\ Vol.\
  \bibinfo {volume} {5412},\ \bibinfo {editor} {edited by\ \bibinfo {editor}
  {\bibfnamefont {G.~W.}\ \bibnamefont {Kamerman}}\ and\ \bibinfo {editor}
  {\bibfnamefont {G.~W.}\ \bibnamefont {Kamerman}}},\ \bibinfo {organization}
  {International Society for Optics and Photonics}\ (\bibinfo  {publisher}
  {SPIE},\ \bibinfo {year} {2004})\ pp.\ \bibinfo {pages} {264 --
  272}\BibitemShut {NoStop}%
\bibitem [{\citenamefont {{Hu}}\ \emph {et~al.}(2017)\citenamefont {{Hu}},
  \citenamefont {{Zhao}}, \citenamefont {{Ye}}, \citenamefont {{Gao}},
  \citenamefont {{Zhao}},\ and\ \citenamefont {{Zhou}}}]{khu_ieeesj2017}%
  \BibitemOpen
  \bibfield  {author} {\bibinfo {author} {\bibfnamefont {K.}~\bibnamefont
  {{Hu}}}, \bibinfo {author} {\bibfnamefont {Y.}~\bibnamefont {{Zhao}}},
  \bibinfo {author} {\bibfnamefont {M.}~\bibnamefont {{Ye}}}, \bibinfo {author}
  {\bibfnamefont {J.}~\bibnamefont {{Gao}}}, \bibinfo {author} {\bibfnamefont
  {G.}~\bibnamefont {{Zhao}}}, \ and\ \bibinfo {author} {\bibfnamefont
  {G.}~\bibnamefont {{Zhou}}},\ }\href {\doibase 10.1109/JSEN.2017.2724064}
  {\bibfield  {journal} {\bibinfo  {journal} {IEEE Sensors Journal}\ }\textbf
  {\bibinfo {volume} {17}},\ \bibinfo {pages} {5547} (\bibinfo {year}
  {2017})}\BibitemShut {NoStop}%
\bibitem [{\citenamefont {Poulton}\ \emph {et~al.}(2017)\citenamefont
  {Poulton}, \citenamefont {Yaacobi}, \citenamefont {Cole}, \citenamefont
  {Byrd}, \citenamefont {Raval}, \citenamefont {Vermeulen},\ and\ \citenamefont
  {Watts}}]{cvpoulton_ol2017}%
  \BibitemOpen
  \bibfield  {author} {\bibinfo {author} {\bibfnamefont {C.~V.}\ \bibnamefont
  {Poulton}}, \bibinfo {author} {\bibfnamefont {A.}~\bibnamefont {Yaacobi}},
  \bibinfo {author} {\bibfnamefont {D.~B.}\ \bibnamefont {Cole}}, \bibinfo
  {author} {\bibfnamefont {M.~J.}\ \bibnamefont {Byrd}}, \bibinfo {author}
  {\bibfnamefont {M.}~\bibnamefont {Raval}}, \bibinfo {author} {\bibfnamefont
  {D.}~\bibnamefont {Vermeulen}}, \ and\ \bibinfo {author} {\bibfnamefont
  {M.~R.}\ \bibnamefont {Watts}},\ }\href {\doibase 10.1364/OL.42.004091}
  {\bibfield  {journal} {\bibinfo  {journal} {Opt. Lett.}\ }\textbf {\bibinfo
  {volume} {42}},\ \bibinfo {pages} {4091} (\bibinfo {year}
  {2017})}\BibitemShut {NoStop}%
\bibitem [{\citenamefont {Miller}\ \emph {et~al.}(2018)\citenamefont {Miller},
  \citenamefont {Phare}, \citenamefont {Chang}, \citenamefont {Ji},
  \citenamefont {Gordillo}, \citenamefont {Mohanty}, \citenamefont {Roberts},
  \citenamefont {Shin}, \citenamefont {Stern}, \citenamefont {Zadka},\ and\
  \citenamefont {Lipson}}]{samiller_cleo2018}%
  \BibitemOpen
  \bibfield  {author} {\bibinfo {author} {\bibfnamefont {S.~A.}\ \bibnamefont
  {Miller}}, \bibinfo {author} {\bibfnamefont {C.~T.}\ \bibnamefont {Phare}},
  \bibinfo {author} {\bibfnamefont {Y.-C.}\ \bibnamefont {Chang}}, \bibinfo
  {author} {\bibfnamefont {X.}~\bibnamefont {Ji}}, \bibinfo {author}
  {\bibfnamefont {O.~A.~J.}\ \bibnamefont {Gordillo}}, \bibinfo {author}
  {\bibfnamefont {A.}~\bibnamefont {Mohanty}}, \bibinfo {author} {\bibfnamefont
  {S.~P.}\ \bibnamefont {Roberts}}, \bibinfo {author} {\bibfnamefont {M.~C.}\
  \bibnamefont {Shin}}, \bibinfo {author} {\bibfnamefont {B.}~\bibnamefont
  {Stern}}, \bibinfo {author} {\bibfnamefont {M.}~\bibnamefont {Zadka}}, \ and\
  \bibinfo {author} {\bibfnamefont {M.}~\bibnamefont {Lipson}},\ }in\ \href
  {\doibase 10.1364/CLEO_AT.2018.JTh5C.2} {\emph {\bibinfo {booktitle}
  {Conference on Lasers and Electro-Optics}}}\ (\bibinfo  {publisher} {Optical
  Society of America},\ \bibinfo {year} {2018})\ p.\ \bibinfo {pages}
  {JTh5C.2}\BibitemShut {NoStop}%
\bibitem [{\citenamefont {{Poulton}}\ \emph {et~al.}(2019)\citenamefont
  {{Poulton}}, \citenamefont {{Byrd}}, \citenamefont {{Russo}}, \citenamefont
  {{Timurdogan}}, \citenamefont {{Khandaker}}, \citenamefont {{Vermeulen}},\
  and\ \citenamefont {{Watts}}}]{cvpoulton_ieeeqe2019}%
  \BibitemOpen
  \bibfield  {author} {\bibinfo {author} {\bibfnamefont {C.~V.}\ \bibnamefont
  {{Poulton}}}, \bibinfo {author} {\bibfnamefont {M.~J.}\ \bibnamefont
  {{Byrd}}}, \bibinfo {author} {\bibfnamefont {P.}~\bibnamefont {{Russo}}},
  \bibinfo {author} {\bibfnamefont {E.}~\bibnamefont {{Timurdogan}}}, \bibinfo
  {author} {\bibfnamefont {M.}~\bibnamefont {{Khandaker}}}, \bibinfo {author}
  {\bibfnamefont {D.}~\bibnamefont {{Vermeulen}}}, \ and\ \bibinfo {author}
  {\bibfnamefont {M.~R.}\ \bibnamefont {{Watts}}},\ }\href {\doibase
  10.1109/JSTQE.2019.2908555} {\bibfield  {journal} {\bibinfo  {journal} {IEEE
  Journal of Selected Topics in Quantum Electronics}\ }\textbf {\bibinfo
  {volume} {25}},\ \bibinfo {pages} {1} (\bibinfo {year} {2019})}\BibitemShut
  {NoStop}%
\bibitem [{\citenamefont {Wang}\ \emph {et~al.}(2015)\citenamefont {Wang},
  \citenamefont {Sun}, \citenamefont {Shou}, \citenamefont {Wang},
  \citenamefont {Wu}, \citenamefont {Chong}, \citenamefont {Liu},\ and\
  \citenamefont {Sun}}]{jwang_jirs2015}%
  \BibitemOpen
  \bibfield  {author} {\bibinfo {author} {\bibfnamefont {J.}~\bibnamefont
  {Wang}}, \bibinfo {author} {\bibfnamefont {W.}~\bibnamefont {Sun}}, \bibinfo
  {author} {\bibfnamefont {W.}~\bibnamefont {Shou}}, \bibinfo {author}
  {\bibfnamefont {X.}~\bibnamefont {Wang}}, \bibinfo {author} {\bibfnamefont
  {C.}~\bibnamefont {Wu}}, \bibinfo {author} {\bibfnamefont {H.-Y.}\
  \bibnamefont {Chong}}, \bibinfo {author} {\bibfnamefont {Y.}~\bibnamefont
  {Liu}}, \ and\ \bibinfo {author} {\bibfnamefont {C.}~\bibnamefont {Sun}},\
  }\href {\doibase 10.1007/s10846-014-0116-8} {\bibfield  {journal} {\bibinfo
  {journal} {Journal of Intelligent \& Robotic Systems}\ }\textbf {\bibinfo
  {volume} {79}},\ \bibinfo {pages} {417} (\bibinfo {year} {2015})}\BibitemShut
  {NoStop}%
\bibitem [{\citenamefont {Kasturi}\ \emph {et~al.}(2016)\citenamefont
  {Kasturi}, \citenamefont {Milanovic}, \citenamefont {Atwood},\ and\
  \citenamefont {Yang}}]{akasturi_lrta2016}%
  \BibitemOpen
  \bibfield  {author} {\bibinfo {author} {\bibfnamefont {A.}~\bibnamefont
  {Kasturi}}, \bibinfo {author} {\bibfnamefont {V.}~\bibnamefont {Milanovic}},
  \bibinfo {author} {\bibfnamefont {B.~H.}\ \bibnamefont {Atwood}}, \ and\
  \bibinfo {author} {\bibfnamefont {J.}~\bibnamefont {Yang}},\ }in\ \href
  {\doibase 10.1117/12.2224285} {\emph {\bibinfo {booktitle} {Laser Radar
  Technology and Applications {XXI}}}},\ Vol.\ \bibinfo {volume} {9832},\
  \bibinfo {editor} {edited by\ \bibinfo {editor} {\bibfnamefont {M.~D.}\
  \bibnamefont {Turner}}\ and\ \bibinfo {editor} {\bibfnamefont {G.~W.}\
  \bibnamefont {Kamerman}}},\ \bibinfo {organization} {International Society
  for Optics and Photonics}\ (\bibinfo  {publisher} {SPIE},\ \bibinfo {year}
  {2016})\ pp.\ \bibinfo {pages} {206 -- 215}\BibitemShut {NoStop}%
\bibitem [{\citenamefont {{Griffiths}}(1990)}]{hdgriffiths_ecej1990}%
  \BibitemOpen
  \bibfield  {author} {\bibinfo {author} {\bibfnamefont {H.~D.}\ \bibnamefont
  {{Griffiths}}},\ }\href {\doibase 10.1049/ecej:19900043} {\bibfield
  {journal} {\bibinfo  {journal} {Electronics Communication Engineering
  Journal}\ }\textbf {\bibinfo {volume} {2}},\ \bibinfo {pages} {185} (\bibinfo
  {year} {1990})}\BibitemShut {NoStop}%
\bibitem [{\citenamefont {Riemensberger}\ \emph {et~al.}(2020)\citenamefont
  {Riemensberger}, \citenamefont {Lukashchuk}, \citenamefont {Karpov},
  \citenamefont {Weng}, \citenamefont {Lucas}, \citenamefont {Liu},\ and\
  \citenamefont {Kippenberg}}]{jriemensberger_nature2020}%
  \BibitemOpen
  \bibfield  {author} {\bibinfo {author} {\bibfnamefont {J.}~\bibnamefont
  {Riemensberger}}, \bibinfo {author} {\bibfnamefont {A.}~\bibnamefont
  {Lukashchuk}}, \bibinfo {author} {\bibfnamefont {M.}~\bibnamefont {Karpov}},
  \bibinfo {author} {\bibfnamefont {W.}~\bibnamefont {Weng}}, \bibinfo {author}
  {\bibfnamefont {E.}~\bibnamefont {Lucas}}, \bibinfo {author} {\bibfnamefont
  {J.}~\bibnamefont {Liu}}, \ and\ \bibinfo {author} {\bibfnamefont {T.~J.}\
  \bibnamefont {Kippenberg}},\ }\href {\doibase 10.1038/s41586-020-2239-3}
  {\bibfield  {journal} {\bibinfo  {journal} {Nature}\ }\textbf {\bibinfo
  {volume} {581}},\ \bibinfo {pages} {164} (\bibinfo {year}
  {2020})}\BibitemShut {NoStop}%
\bibitem [{\citenamefont {{Thurn}}\ \emph {et~al.}(2013)\citenamefont
  {{Thurn}}, \citenamefont {{Ebelt}},\ and\ \citenamefont
  {{Vossiek}}}]{kthurn_ieeemtt2013}%
  \BibitemOpen
  \bibfield  {author} {\bibinfo {author} {\bibfnamefont {K.}~\bibnamefont
  {{Thurn}}}, \bibinfo {author} {\bibfnamefont {R.}~\bibnamefont {{Ebelt}}}, \
  and\ \bibinfo {author} {\bibfnamefont {M.}~\bibnamefont {{Vossiek}}},\ }in\
  \href {\doibase 10.1109/MWSYM.2013.6697654} {\emph {\bibinfo {booktitle}
  {2013 IEEE MTT-S International Microwave Symposium Digest (MTT)}}}\ (\bibinfo
  {year} {2013})\ pp.\ \bibinfo {pages} {1--3}\BibitemShut {NoStop}%
\bibitem [{\citenamefont {Tsang}\ \emph {et~al.}(2002)\citenamefont {Tsang},
  \citenamefont {Wong}, \citenamefont {Liang}, \citenamefont {Day},
  \citenamefont {Roberts}, \citenamefont {Harpin}, \citenamefont {Drake},\ and\
  \citenamefont {Asghari}}]{hktsang_apl2002}%
  \BibitemOpen
  \bibfield  {author} {\bibinfo {author} {\bibfnamefont {H.~K.}\ \bibnamefont
  {Tsang}}, \bibinfo {author} {\bibfnamefont {C.~S.}\ \bibnamefont {Wong}},
  \bibinfo {author} {\bibfnamefont {T.~K.}\ \bibnamefont {Liang}}, \bibinfo
  {author} {\bibfnamefont {I.~E.}\ \bibnamefont {Day}}, \bibinfo {author}
  {\bibfnamefont {S.~W.}\ \bibnamefont {Roberts}}, \bibinfo {author}
  {\bibfnamefont {A.}~\bibnamefont {Harpin}}, \bibinfo {author} {\bibfnamefont
  {J.}~\bibnamefont {Drake}}, \ and\ \bibinfo {author} {\bibfnamefont
  {M.}~\bibnamefont {Asghari}},\ }\href {\doibase 10.1063/1.1435801} {\bibfield
   {journal} {\bibinfo  {journal} {Applied Physics Letters}\ }\textbf {\bibinfo
  {volume} {80}},\ \bibinfo {pages} {416} (\bibinfo {year} {2002})}\BibitemShut
  {NoStop}%
\bibitem [{\citenamefont {Rong}\ \emph {et~al.}(2005)\citenamefont {Rong},
  \citenamefont {Liu}, \citenamefont {Jones}, \citenamefont {Cohen},
  \citenamefont {Hak}, \citenamefont {Nicolaescu}, \citenamefont {Fang},\ and\
  \citenamefont {Paniccia}}]{hrong_nature2005}%
  \BibitemOpen
  \bibfield  {author} {\bibinfo {author} {\bibfnamefont {H.}~\bibnamefont
  {Rong}}, \bibinfo {author} {\bibfnamefont {A.}~\bibnamefont {Liu}}, \bibinfo
  {author} {\bibfnamefont {R.}~\bibnamefont {Jones}}, \bibinfo {author}
  {\bibfnamefont {O.}~\bibnamefont {Cohen}}, \bibinfo {author} {\bibfnamefont
  {D.}~\bibnamefont {Hak}}, \bibinfo {author} {\bibfnamefont {R.}~\bibnamefont
  {Nicolaescu}}, \bibinfo {author} {\bibfnamefont {A.}~\bibnamefont {Fang}}, \
  and\ \bibinfo {author} {\bibfnamefont {M.}~\bibnamefont {Paniccia}},\ }\href
  {\doibase 10.1038/nature03273} {\bibfield  {journal} {\bibinfo  {journal}
  {Nature}\ }\textbf {\bibinfo {volume} {433}},\ \bibinfo {pages} {292}
  (\bibinfo {year} {2005})}\BibitemShut {NoStop}%
\bibitem [{\citenamefont {{Giewont}}\ \emph {et~al.}(2019)\citenamefont
  {{Giewont}}, \citenamefont {{Nummy}}, \citenamefont {{Anderson}},
  \citenamefont {{Ayala}}, \citenamefont {{Barwicz}}, \citenamefont {{Bian}},
  \citenamefont {{Dezfulian}}, \citenamefont {{Gill}}, \citenamefont
  {{Houghton}}, \citenamefont {{Hu}}, \citenamefont {{Peng}}, \citenamefont
  {{Rakowski}}, \citenamefont {{Rauch}}, \citenamefont {{Rosenberg}},
  \citenamefont {{Sahin}}, \citenamefont {{Stobert}},\ and\ \citenamefont
  {{Stricker}}}]{kgiewont_ieeeqe2019}%
  \BibitemOpen
  \bibfield  {author} {\bibinfo {author} {\bibfnamefont {K.}~\bibnamefont
  {{Giewont}}}, \bibinfo {author} {\bibfnamefont {K.}~\bibnamefont {{Nummy}}},
  \bibinfo {author} {\bibfnamefont {F.~A.}\ \bibnamefont {{Anderson}}},
  \bibinfo {author} {\bibfnamefont {J.}~\bibnamefont {{Ayala}}}, \bibinfo
  {author} {\bibfnamefont {T.}~\bibnamefont {{Barwicz}}}, \bibinfo {author}
  {\bibfnamefont {Y.}~\bibnamefont {{Bian}}}, \bibinfo {author} {\bibfnamefont
  {K.~K.}\ \bibnamefont {{Dezfulian}}}, \bibinfo {author} {\bibfnamefont
  {D.~M.}\ \bibnamefont {{Gill}}}, \bibinfo {author} {\bibfnamefont
  {T.}~\bibnamefont {{Houghton}}}, \bibinfo {author} {\bibfnamefont
  {S.}~\bibnamefont {{Hu}}}, \bibinfo {author} {\bibfnamefont {B.}~\bibnamefont
  {{Peng}}}, \bibinfo {author} {\bibfnamefont {M.}~\bibnamefont {{Rakowski}}},
  \bibinfo {author} {\bibfnamefont {S.}~\bibnamefont {{Rauch}}}, \bibinfo
  {author} {\bibfnamefont {J.~C.}\ \bibnamefont {{Rosenberg}}}, \bibinfo
  {author} {\bibfnamefont {A.}~\bibnamefont {{Sahin}}}, \bibinfo {author}
  {\bibfnamefont {I.}~\bibnamefont {{Stobert}}}, \ and\ \bibinfo {author}
  {\bibfnamefont {A.}~\bibnamefont {{Stricker}}},\ }\href {\doibase
  10.1109/JSTQE.2019.2908790} {\bibfield  {journal} {\bibinfo  {journal} {IEEE
  Journal of Selected Topics in Quantum Electronics}\ }\textbf {\bibinfo
  {volume} {25}},\ \bibinfo {pages} {1} (\bibinfo {year} {2019})}\BibitemShut
  {NoStop}%
\bibitem [{\citenamefont {Shen}\ \emph {et~al.}(2017)\citenamefont {Shen},
  \citenamefont {Harris}, \citenamefont {Skirlo}, \citenamefont {Prabhu},
  \citenamefont {Baehr-Jones}, \citenamefont {Hochberg}, \citenamefont {Sun},
  \citenamefont {Zhao}, \citenamefont {Larochelle}, \citenamefont {Englund},\
  and\ \citenamefont {{Solja\v{c}i\'{c}}}}]{yshen_nphoton2017}%
  \BibitemOpen
  \bibfield  {author} {\bibinfo {author} {\bibfnamefont {Y.}~\bibnamefont
  {Shen}}, \bibinfo {author} {\bibfnamefont {N.~C.}\ \bibnamefont {Harris}},
  \bibinfo {author} {\bibfnamefont {S.}~\bibnamefont {Skirlo}}, \bibinfo
  {author} {\bibfnamefont {M.}~\bibnamefont {Prabhu}}, \bibinfo {author}
  {\bibfnamefont {T.}~\bibnamefont {Baehr-Jones}}, \bibinfo {author}
  {\bibfnamefont {M.}~\bibnamefont {Hochberg}}, \bibinfo {author}
  {\bibfnamefont {X.}~\bibnamefont {Sun}}, \bibinfo {author} {\bibfnamefont
  {S.}~\bibnamefont {Zhao}}, \bibinfo {author} {\bibfnamefont {H.}~\bibnamefont
  {Larochelle}}, \bibinfo {author} {\bibfnamefont {D.}~\bibnamefont {Englund}},
  \ and\ \bibinfo {author} {\bibfnamefont {M.}~\bibnamefont
  {{Solja\v{c}i\'{c}}}},\ }\href {\doibase 10.1038/nphoton.2017.93} {\bibfield
  {journal} {\bibinfo  {journal} {Nature Photonics}\ }\textbf {\bibinfo
  {volume} {11}},\ \bibinfo {pages} {441} (\bibinfo {year} {2017})}\BibitemShut
  {NoStop}%
\bibitem [{\citenamefont {{Razavi}}(2019)}]{razavi_sscm2019}%
  \BibitemOpen
  \bibfield  {author} {\bibinfo {author} {\bibfnamefont {B.}~\bibnamefont
  {{Razavi}}},\ }\href@noop {} {\bibfield  {journal} {\bibinfo  {journal}
  {Solid-State Circuits Magazine}\ }\textbf {\bibinfo {volume} {11}},\ \bibinfo
  {pages} {10} (\bibinfo {year} {2019})}\BibitemShut {NoStop}%
\bibitem [{\citenamefont {Wang}(1982)}]{jywang_ao1982}%
  \BibitemOpen
  \bibfield  {author} {\bibinfo {author} {\bibfnamefont {J.~Y.}\ \bibnamefont
  {Wang}},\ }\href {\doibase 10.1364/AO.21.000464} {\bibfield  {journal}
  {\bibinfo  {journal} {Appl. Opt.}\ }\textbf {\bibinfo {volume} {21}},\
  \bibinfo {pages} {464} (\bibinfo {year} {1982})}\BibitemShut {NoStop}%
\bibitem [{\citenamefont {Rebolj}\ \emph {et~al.}(2017)\citenamefont {Rebolj},
  \citenamefont {Pu{\v{c}}ko}, \citenamefont {Babi{\v{c}}}, \citenamefont
  {Bizjak},\ and\ \citenamefont {Mongus}}]{drebolj_ac2017}%
  \BibitemOpen
  \bibfield  {author} {\bibinfo {author} {\bibfnamefont {D.}~\bibnamefont
  {Rebolj}}, \bibinfo {author} {\bibfnamefont {Z.}~\bibnamefont {Pu{\v{c}}ko}},
  \bibinfo {author} {\bibfnamefont {N.~{\v{C}u\v{s}}.}\ \bibnamefont
  {Babi{\v{c}}}}, \bibinfo {author} {\bibfnamefont {M.}~\bibnamefont {Bizjak}},
  \ and\ \bibinfo {author} {\bibfnamefont {D.}~\bibnamefont {Mongus}},\ }\href
  {\doibase 10.1016/j.autcon.2017.09.021} {\bibfield  {journal} {\bibinfo
  {journal} {Automation in Construction}\ }\textbf {\bibinfo {volume} {84}},\
  \bibinfo {pages} {323 } (\bibinfo {year} {2017})}\BibitemShut {NoStop}%
\bibitem [{\citenamefont {{Samani}}\ \emph {et~al.}(2019)\citenamefont
  {{Samani}}, \citenamefont {{El-Fiky}}, \citenamefont {{Morsy-Osman}},
  \citenamefont {{Li}}, \citenamefont {{Patel}}, \citenamefont {{Hoang}},
  \citenamefont {{Jacques}}, \citenamefont {{Chagnon}}, \citenamefont
  {{Abadía}},\ and\ \citenamefont {{Plant}}}]{asamani_jlwt2019}%
  \BibitemOpen
  \bibfield  {author} {\bibinfo {author} {\bibfnamefont {A.}~\bibnamefont
  {{Samani}}}, \bibinfo {author} {\bibfnamefont {E.}~\bibnamefont {{El-Fiky}}},
  \bibinfo {author} {\bibfnamefont {M.}~\bibnamefont {{Morsy-Osman}}}, \bibinfo
  {author} {\bibfnamefont {R.}~\bibnamefont {{Li}}}, \bibinfo {author}
  {\bibfnamefont {D.}~\bibnamefont {{Patel}}}, \bibinfo {author} {\bibfnamefont
  {T.}~\bibnamefont {{Hoang}}}, \bibinfo {author} {\bibfnamefont
  {M.}~\bibnamefont {{Jacques}}}, \bibinfo {author} {\bibfnamefont
  {M.}~\bibnamefont {{Chagnon}}}, \bibinfo {author} {\bibfnamefont
  {N.}~\bibnamefont {{Abadía}}}, \ and\ \bibinfo {author} {\bibfnamefont
  {D.~V.}\ \bibnamefont {{Plant}}},\ }\href {\doibase 10.1109/JLT.2019.2908655}
  {\bibfield  {journal} {\bibinfo  {journal} {Journal of Lightwave Technology}\
  }\textbf {\bibinfo {volume} {37}},\ \bibinfo {pages} {2989} (\bibinfo {year}
  {2019})}\BibitemShut {NoStop}%
\bibitem [{\citenamefont {Stove}(1992)}]{agstove_ipf1992}%
  \BibitemOpen
  \bibfield  {author} {\bibinfo {author} {\bibfnamefont {A.}~\bibnamefont
  {Stove}},\ }\href {\doibase 10.1049/ip-f-2.1992.0048} {\bibfield  {journal}
  {\bibinfo  {journal} {IEE Proceedings F (Radar and Signal Processing)}\
  }\textbf {\bibinfo {volume} {139}},\ \bibinfo {pages} {343} (\bibinfo {year}
  {1992})}\BibitemShut {NoStop}%
\bibitem [{\citenamefont {{Winkler}}(2007)}]{vwinkler_erc2007}%
  \BibitemOpen
  \bibfield  {author} {\bibinfo {author} {\bibfnamefont {V.}~\bibnamefont
  {{Winkler}}},\ }in\ \href {\doibase 10.1109/EURAD.2007.4404963} {\emph
  {\bibinfo {booktitle} {2007 European Radar Conference}}}\ (\bibinfo {year}
  {2007})\ pp.\ \bibinfo {pages} {166--169}\BibitemShut {NoStop}%
\bibitem [{\citenamefont {Chen}\ and\ \citenamefont
  {Rao}(1968)}]{hschen_jpd1968}%
  \BibitemOpen
  \bibfield  {author} {\bibinfo {author} {\bibfnamefont {H.-S.}\ \bibnamefont
  {Chen}}\ and\ \bibinfo {author} {\bibfnamefont {C.~R.~N.}\ \bibnamefont
  {Rao}},\ }\href {\doibase 10.1088/0022-3727/1/9/312} {\bibfield  {journal}
  {\bibinfo  {journal} {Journal of Physics D: Applied Physics}\ }\textbf
  {\bibinfo {volume} {1}},\ \bibinfo {pages} {1191} (\bibinfo {year}
  {1968})}\BibitemShut {NoStop}%
\bibitem [{\citenamefont {{Sheng}}\ \emph {et~al.}(2012)\citenamefont
  {{Sheng}}, \citenamefont {{Wang}}, \citenamefont {{Qiu}}, \citenamefont
  {{Li}}, \citenamefont {{Pang}}, \citenamefont {{Wu}}, \citenamefont {{Wang}},
  \citenamefont {{Zou}},\ and\ \citenamefont {{Gan}}}]{zsheng_ieeepj2012}%
  \BibitemOpen
  \bibfield  {author} {\bibinfo {author} {\bibfnamefont {Z.}~\bibnamefont
  {{Sheng}}}, \bibinfo {author} {\bibfnamefont {Z.}~\bibnamefont {{Wang}}},
  \bibinfo {author} {\bibfnamefont {C.}~\bibnamefont {{Qiu}}}, \bibinfo
  {author} {\bibfnamefont {L.}~\bibnamefont {{Li}}}, \bibinfo {author}
  {\bibfnamefont {A.}~\bibnamefont {{Pang}}}, \bibinfo {author} {\bibfnamefont
  {A.}~\bibnamefont {{Wu}}}, \bibinfo {author} {\bibfnamefont {X.}~\bibnamefont
  {{Wang}}}, \bibinfo {author} {\bibfnamefont {S.}~\bibnamefont {{Zou}}}, \
  and\ \bibinfo {author} {\bibfnamefont {F.}~\bibnamefont {{Gan}}},\ }\href
  {\doibase 10.1109/JPHOT.2012.2230320} {\bibfield  {journal} {\bibinfo
  {journal} {IEEE Photonics Journal}\ }\textbf {\bibinfo {volume} {4}},\
  \bibinfo {pages} {2272} (\bibinfo {year} {2012})}\BibitemShut {NoStop}%
\bibitem [{\citenamefont {Harris}\ \emph {et~al.}(2014)\citenamefont {Harris},
  \citenamefont {Ma}, \citenamefont {Mower}, \citenamefont {Baehr-Jones},
  \citenamefont {Englund}, \citenamefont {Hochberg},\ and\ \citenamefont
  {Galland}}]{ncharris_oe2014}%
  \BibitemOpen
  \bibfield  {author} {\bibinfo {author} {\bibfnamefont {N.~C.}\ \bibnamefont
  {Harris}}, \bibinfo {author} {\bibfnamefont {Y.}~\bibnamefont {Ma}}, \bibinfo
  {author} {\bibfnamefont {J.}~\bibnamefont {Mower}}, \bibinfo {author}
  {\bibfnamefont {T.}~\bibnamefont {Baehr-Jones}}, \bibinfo {author}
  {\bibfnamefont {D.}~\bibnamefont {Englund}}, \bibinfo {author} {\bibfnamefont
  {M.}~\bibnamefont {Hochberg}}, \ and\ \bibinfo {author} {\bibfnamefont
  {C.}~\bibnamefont {Galland}},\ }\href {\doibase 10.1364/OE.22.010487}
  {\bibfield  {journal} {\bibinfo  {journal} {Opt. Express}\ }\textbf {\bibinfo
  {volume} {22}},\ \bibinfo {pages} {10487} (\bibinfo {year}
  {2014})}\BibitemShut {NoStop}%
\bibitem [{\citenamefont {Mendez-Astudillo}\ \emph {et~al.}(2019)\citenamefont
  {Mendez-Astudillo}, \citenamefont {Okamoto}, \citenamefont {Ito},\ and\
  \citenamefont {Kita}}]{smendezastudillo_oe2019}%
  \BibitemOpen
  \bibfield  {author} {\bibinfo {author} {\bibfnamefont {M.}~\bibnamefont
  {Mendez-Astudillo}}, \bibinfo {author} {\bibfnamefont {M.}~\bibnamefont
  {Okamoto}}, \bibinfo {author} {\bibfnamefont {Y.}~\bibnamefont {Ito}}, \ and\
  \bibinfo {author} {\bibfnamefont {T.}~\bibnamefont {Kita}},\ }\href {\doibase
  10.1364/OE.27.000899} {\bibfield  {journal} {\bibinfo  {journal} {Opt.
  Express}\ }\textbf {\bibinfo {volume} {27}},\ \bibinfo {pages} {899}
  (\bibinfo {year} {2019})}\BibitemShut {NoStop}%
\bibitem [{\citenamefont {{Ahmed}}\ \emph {et~al.}(2019)\citenamefont
  {{Ahmed}}, \citenamefont {{Huynh}}, \citenamefont {{Williams}}, \citenamefont
  {{Wang}}, \citenamefont {{Hanumolu}},\ and\ \citenamefont
  {{Rylyakov}}}]{ahmed_jssc2019}%
  \BibitemOpen
  \bibfield  {author} {\bibinfo {author} {\bibfnamefont {M.~G.}\ \bibnamefont
  {{Ahmed}}}, \bibinfo {author} {\bibfnamefont {T.~N.}\ \bibnamefont
  {{Huynh}}}, \bibinfo {author} {\bibfnamefont {C.}~\bibnamefont {{Williams}}},
  \bibinfo {author} {\bibfnamefont {Y.}~\bibnamefont {{Wang}}}, \bibinfo
  {author} {\bibfnamefont {P.~K.}\ \bibnamefont {{Hanumolu}}}, \ and\ \bibinfo
  {author} {\bibfnamefont {A.}~\bibnamefont {{Rylyakov}}},\ }\href {\doibase
  10.1109/JSSC.2018.2882265} {\bibfield  {journal} {\bibinfo  {journal} {IEEE
  Journal of Solid-State Circuits}\ }\textbf {\bibinfo {volume} {54}},\
  \bibinfo {pages} {834} (\bibinfo {year} {2019})}\BibitemShut {NoStop}%
\bibitem [{\citenamefont {Shahdoost}\ \emph {et~al.}(2016)\citenamefont
  {Shahdoost}, \citenamefont {Medi},\ and\ \citenamefont
  {Saniei}}]{shahdoost_ana2016}%
  \BibitemOpen
  \bibfield  {author} {\bibinfo {author} {\bibfnamefont {S.}~\bibnamefont
  {Shahdoost}}, \bibinfo {author} {\bibfnamefont {A.}~\bibnamefont {Medi}}, \
  and\ \bibinfo {author} {\bibfnamefont {N.}~\bibnamefont {Saniei}},\ }\href
  {\doibase 10.1007/s10470-015-0669-x} {\bibfield  {journal} {\bibinfo
  {journal} {Analog Integrated Circuits and Signal Processing}\ }\textbf
  {\bibinfo {volume} {86}},\ \bibinfo {pages} {233} (\bibinfo {year}
  {2016})}\BibitemShut {NoStop}%
\bibitem [{\citenamefont {{Mohan}}\ \emph {et~al.}(2000)\citenamefont
  {{Mohan}}, \citenamefont {{Hershenson}}, \citenamefont {{Boyd}},\ and\
  \citenamefont {{Lee}}}]{mohan_jssc2000}%
  \BibitemOpen
  \bibfield  {author} {\bibinfo {author} {\bibfnamefont {S.~S.}\ \bibnamefont
  {{Mohan}}}, \bibinfo {author} {\bibfnamefont {M.~D.~M.}\ \bibnamefont
  {{Hershenson}}}, \bibinfo {author} {\bibfnamefont {S.~P.}\ \bibnamefont
  {{Boyd}}}, \ and\ \bibinfo {author} {\bibfnamefont {T.~H.}\ \bibnamefont
  {{Lee}}},\ }\href {\doibase 10.1109/4.826816} {\bibfield  {journal} {\bibinfo
   {journal} {IEEE Journal of Solid-State Circuits}\ }\textbf {\bibinfo
  {volume} {35}},\ \bibinfo {pages} {346} (\bibinfo {year} {2000})}\BibitemShut
  {NoStop}%
\bibitem [{\citenamefont {{Razavi}}(2000)}]{razavi_isscc2000}%
  \BibitemOpen
  \bibfield  {author} {\bibinfo {author} {\bibfnamefont {B.}~\bibnamefont
  {{Razavi}}},\ }in\ \href {\doibase 10.1109/ISSCC.2000.839732} {\emph
  {\bibinfo {booktitle} {2000 IEEE International Solid-State Circuits
  Conference. Digest of Technical Papers (Cat. No.00CH37056)}}}\ (\bibinfo
  {year} {2000})\ pp.\ \bibinfo {pages} {162--163}\BibitemShut {NoStop}%
\bibitem [{\citenamefont {Romanova}\ and\ \citenamefont
  {Barzdenas}(2019)}]{romanova_elec2019}%
  \BibitemOpen
  \bibfield  {author} {\bibinfo {author} {\bibfnamefont {A.}~\bibnamefont
  {Romanova}}\ and\ \bibinfo {author} {\bibfnamefont {V.}~\bibnamefont
  {Barzdenas}},\ }\href {\doibase 10.3390/electronics8101073} {\bibfield
  {journal} {\bibinfo  {journal} {Electronics}\ }\textbf {\bibinfo {volume}
  {8}},\ \bibinfo {pages} {1073} (\bibinfo {year} {2019})}\BibitemShut
  {NoStop}%
\bibitem [{\citenamefont {Sun}\ \emph {et~al.}(2013)\citenamefont {Sun},
  \citenamefont {Timurdogan}, \citenamefont {Yaacobi}, \citenamefont
  {Hosseini},\ and\ \citenamefont {Watts}}]{jsun_nature2013}%
  \BibitemOpen
  \bibfield  {author} {\bibinfo {author} {\bibfnamefont {J.}~\bibnamefont
  {Sun}}, \bibinfo {author} {\bibfnamefont {E.}~\bibnamefont {Timurdogan}},
  \bibinfo {author} {\bibfnamefont {A.}~\bibnamefont {Yaacobi}}, \bibinfo
  {author} {\bibfnamefont {E.~S.}\ \bibnamefont {Hosseini}}, \ and\ \bibinfo
  {author} {\bibfnamefont {M.~R.}\ \bibnamefont {Watts}},\ }\href {\doibase
  10.1038/nature11727} {\bibfield  {journal} {\bibinfo  {journal} {Nature}\
  }\textbf {\bibinfo {volume} {493}},\ \bibinfo {pages} {195} (\bibinfo {year}
  {2013})}\BibitemShut {NoStop}%
\bibitem [{\citenamefont {Piggott}\ \emph {et~al.}(2020)\citenamefont
  {Piggott}, \citenamefont {Ma}, \citenamefont {Su}, \citenamefont {Ahn},
  \citenamefont {Sapra}, \citenamefont {Vercruysse}, \citenamefont {Netherton},
  \citenamefont {Khope}, \citenamefont {Bowers},\ and\ \citenamefont
  {Vučković}}]{aypiggott_acsphoton2020}%
  \BibitemOpen
  \bibfield  {author} {\bibinfo {author} {\bibfnamefont {A.~Y.}\ \bibnamefont
  {Piggott}}, \bibinfo {author} {\bibfnamefont {E.~Y.}\ \bibnamefont {Ma}},
  \bibinfo {author} {\bibfnamefont {L.}~\bibnamefont {Su}}, \bibinfo {author}
  {\bibfnamefont {G.~H.}\ \bibnamefont {Ahn}}, \bibinfo {author} {\bibfnamefont
  {N.~V.}\ \bibnamefont {Sapra}}, \bibinfo {author} {\bibfnamefont
  {D.}~\bibnamefont {Vercruysse}}, \bibinfo {author} {\bibfnamefont {A.~M.}\
  \bibnamefont {Netherton}}, \bibinfo {author} {\bibfnamefont {A.~S.~P.}\
  \bibnamefont {Khope}}, \bibinfo {author} {\bibfnamefont {J.~E.}\ \bibnamefont
  {Bowers}}, \ and\ \bibinfo {author} {\bibfnamefont {J.}~\bibnamefont
  {Vučković}},\ }\href {\doibase 10.1021/acsphotonics.9b01540} {\bibfield
  {journal} {\bibinfo  {journal} {ACS Photonics}\ }\textbf {\bibinfo {volume}
  {7}},\ \bibinfo {pages} {569} (\bibinfo {year} {2020})}\BibitemShut {NoStop}%
\bibitem [{\citenamefont {Sorianello}\ \emph {et~al.}(2008)\citenamefont
  {Sorianello}, \citenamefont {Perna}, \citenamefont {Colace}, \citenamefont
  {Assanto}, \citenamefont {Luan},\ and\ \citenamefont
  {Kimerling}}]{vsorianello_apl2008}%
  \BibitemOpen
  \bibfield  {author} {\bibinfo {author} {\bibfnamefont {V.}~\bibnamefont
  {Sorianello}}, \bibinfo {author} {\bibfnamefont {A.}~\bibnamefont {Perna}},
  \bibinfo {author} {\bibfnamefont {L.}~\bibnamefont {Colace}}, \bibinfo
  {author} {\bibfnamefont {G.}~\bibnamefont {Assanto}}, \bibinfo {author}
  {\bibfnamefont {H.~C.}\ \bibnamefont {Luan}}, \ and\ \bibinfo {author}
  {\bibfnamefont {L.~C.}\ \bibnamefont {Kimerling}},\ }\href {\doibase
  10.1063/1.2987999} {\bibfield  {journal} {\bibinfo  {journal} {Applied
  Physics Letters}\ }\textbf {\bibinfo {volume} {93}},\ \bibinfo {pages}
  {111115} (\bibinfo {year} {2008})}\BibitemShut {NoStop}%
\end{thebibliography}%

%%%%%%%%%% Merge with supplemental materials %%%%%%%%%%
\clearpage
\onecolumngrid

\begin{center}
\textbf{\large Methods}
\end{center}

\twocolumngrid
%%%%%%%%%% Merge with supplemental materials %%%%%%%%%%
\subsection{Design and fabrication.}
The demonstration chips used as transmitter and receiver FPAs were fabricated using GlobalFoundries' CMS90WG 300 mm silicon photonics process\cite{kgiewont_ieeeqe2019}, which monolithically integrates photonic devices with $90~\mathrm{nm}$ silicon-on-insulator (SOI) RF CMOS electronics. All photonic devices used in the design, with the exception of the directional couplers, were provided in the foundry's standard process development kit (PDK). By doing so the photonic architecture had correct-by-construction device placement and connectivity, verifiable using Mentor Graphics' Calibre Design Rule Checker. The integrated electronics followed a standard design flow using Cadence Virtuoso and Spectre for circuit design and layout, and Mentor Graphics' Calibre for verification of design rules, comparing layout-versus-schematic, and extracting parasitics. The two domains are merged into a single hierarchy enabling connectivity verification at the receiver photodiodes along with design rule verification of closely intertwined photonics and electronics across the chip.

Due to the limited number of dies available from the multi-project wafer shuttle, we were able to fully test the functionality of 5 dies. We did not observe any defects such as dead pixels or inoperative thermo-optic switches across these dies.

\subsection{Optical chirp scheme.}
A linearly chirped optical field $E(t)$ has the form
\begin{align} E(t) &= \exp\left(i 2 \pi f_0 t + i \pi r t^2 \right) \nonumber \\ &= \left[ \cos \pi r t^2 + i\sin \pi r t^2 \right] \exp(i 2 \pi f_0 t), \end{align}
where $f_0$ is the carrier frequency, and $r$ is the chirp ramp rate. Thus, by coherently modulating fixed-frequency light with a microwave chirp of the form $\cos \pi r t^2 + i\sin \pi r t^2 $, we produce a linear chirp in the optical domain.

In our demonstrator system, we use a series of linear up-chirps immediately followed by a series of down-chirps. The mean and difference of the up-chirp and down-chirp beat frequencies allow separate measurement of range and velocity of a target respectively\cite{bbehroozpour_ieeecm2017, hdgriffiths_ecej1990, jriemensberger_nature2020}.

For low-velocity measurements, which includes all the measurements presented here except those in Extended Data Figure \ref{extfig:5_system_perf}(f), we use a single up-chirp and single down-chirp, each with a length of $850~\mathrm{\mu s}$. This chirp scheme is illustrated in Extended Data Figure {\ref{extfig:5_system_perf}}(c). The beat frequencies for each chirp were then extracted using fast-Fourier transforms (FFTs). For fast moving objects with sufficiently large Doppler shifts, however, the beat frequency wraps around zero, resulting in ambiguous measurements. To compensate for this effect, high velocity measurements were performed using a series of fifty $17~\mathrm{\mu s}$ up-chirps, followed by fifty $17~\mathrm{\mu s}$ down-chirps, as shown in Extended Data Figure \ref{extfig:5_system_perf}(d). The multiple chirps were coherently combined using a two-dimensional FFT to extract the beat frequencies, maintaining the same signal-to-noise ratio as the single-chirp case \cite{agstove_ipf1992, vwinkler_erc2007}.

\subsection{Optical setup.}
A narrow-linewidth $(<100~\mathrm{Hz})$ fiber laser (NKT Adjustik) operating at $1550~\mathrm{nm}$ was used as the seed laser for the FMCW ranging system. A linear chirp was applied to the laser light using a silicon photonic IQ modulator fabricated at the University of Southampton, which was driven by a microwave chirp produced by an arbitrary waveform generator (Tektronix AWG70002A). The chirped laser light was amplified by erbium-doped fiber amplifiers (EDFAs) in two stages (a Keopsys CEFA-C-HG-PM followed by an NKT Boostik). The amplified light was then coupled on-chip via single-mode optical fiber V-grooves into two identical demonstration chips, used as a transmitter and receiver respectively. The light emitted by the transmitter FPA was structured using a 32x16 microlens array (PowerPhotonic), which was precisely matched to the receiver array's pixel pattern. This created a structured illumination pattern and minimized any waste of light due to transmit optical power being incident on the gaps between pixels.

To precisely match the fields of view of the transmitter and receiver FPAs, we took advantage of the fact that each receiver grating coupler emits a small amount of LO light due to backreflections from the balanced detectors. An infrared camera was then used to align the patterns of spots produced by the receiver and transmitter FPAs.

The physical aperture of the output lens is 25 mm in diameter. The size of the mode corresponding to a single grating coupler on the receiver focal plane array at this same lens (which can be thought of as the entrance pupil of our system) is slightly elliptical and is 11 mm along the first axis and 16 mm along the second axis. For the long range $75~\mathrm{m}$ measurements, the output lens was adjusted such that the transmit and receive beams were essentially collimated to infinity. For shorter range measurements, the lens position was adjusted to simulate the effects of a smaller aperture.

Due to the use of coherent detection, the receiver detects a single polarization of scattered light. In coherent LiDAR systems, the receiver polarization is typically chosen to be the same as the transmitter polarization\cite{bbehroozpour_ieeecm2017, hdgriffiths_ecej1990, jriemensberger_nature2020} 
as is done here, since most materials preferentially scatter light into the same polarization as the illuminating light\cite{hschen_jpd1968}.

\subsection{On-chip electronics.}
As illustrated in Extended Data Fig. \ref{extfig:2_readout_arch_and_synch}(a-c), there are several levels of on-chip multiplexing in the receiver which allow pixels to be read out in blocks of 8 at a time. The lowest level of multiplexing is achieved by making use of the power switch incorporated into each pixel's TIA: only one pixel per row is activated at a time. The appropriate receiver block is then selected by activating the set of eight buffer amplifiers associated with that block. A final set of differential output amplifiers drives eight off-chip $100~\Omega$ loads. An inter-stage RC filter flattens the frequency response to simplify downstream signal processing.

The receiver bandwidth of $280~\mathrm{MHz}$ in Fig. \ref{fig:3_electrooptic_perf}(a) is the measured bandwidth through the packaged chip. The simulated bandwidth of the on-chip amplifier chain is $750~\mathrm{MHz}$, suggesting a bandwidth limitation in the test setup.

\subsection{Thermo-optic switching tree operation.}
Operation of the thermo-optic switching trees is demonstrated in Extended Data Fig. \ref{extfig:1_switch_trees}. Both the transmit and receive thermo-optic switch trees on our demonstration chip contained integrated photodiodes to monitor the flow of light through the switching trees. To enable digital control and calibration, the monitor photodiodes were directly connected to off-chip TIAs and analog-to-digital converters (Analog Devices AD7091R-8), and the thermo-optic phase shifters were connected to off-chip digital-to-analog converters (Analog Devices AD5391). The switch trees were calibrated one switch at a time by adjusting the control voltage to maximize the optical power in each of the tree outputs. Due to minor thermal cross-talk between the thermo-optic switches, it was necessary to repeat this process for several iterations to converge on an optimal configuration. The average power consumption of a single thermo-optic switch was $40~\mathrm{mW}$. This leads to an average power consumption of approximately $160~\mathrm{mW}$ and $120~\mathrm{mW}$ for the transmitter and receiver switching trees, respectively. The thermo-optic switching trees were found to be very stable, with no recalibration required even after several months of operation in an uncontrolled temperature environment.

\subsection{Transmitter chip losses.}
Optical losses in the switching trees are estimated to be $ 2.1~\mathrm{dB}$ per switch layer, due in large part to sub-optimal $2 \times 2$ couplers. For the transmitter, 4 layers of switches, an additional $ 1~\mathrm{dB}$ of waveguide routing loss, and $ 1.5~\mathrm{dB}$ of grating coupler loss leads to total linear on-chip losses of $ 10.9~\mathrm{dB}$. Under typical transmitter operating conditions, there is an extra $ 2~\mathrm{dB}$ loss from two-photon absorption.

These losses can be significantly reduced in a straightforward manner. For example, by employing existing designs for $2 \times 2$ couplers with a loss of $0.06~\mathrm{dB}$ \cite{zsheng_ieeepj2012}, and thermo-optic phase shifters with a demonstrated loss of $0.23~\mathrm{dB}$ \cite{ncharris_oe2014}, the loss could be reduced to $0.35~\mathrm{dB}$ per switch. Meanwhile, two-photon absorption can be minimized by using wider waveguides or reverse-biased PIN junctions\cite{hrong_nature2005}.

\subsection{Electronic control and signal processing.}
A field-programmable gate array (FPGA) with integrated RF analog-to-digital converters (ADCs) and digital-to-analog converters (DACs)  was used for both system control and signal acquisition (Xilinx Zynq UltraScale+ RFSoC). Thermo-optic switch control and receiver array multiplexing were coordinated by software running on the system's ARM Cortex-A53 processing core. On the signal processing side, the 8 receiver output signals were first digitized in parallel using 8 integrated ADCs, followed by decimation, application of a Hann window, FFTs, and peak detection on a custom digital signal processing (DSP) pipeline. Final data processing and point cloud reconstruction were performed on a personal computer. To precisely measure the beat frequencies, we performed a least-squares fit of the expected lineshape to each peak in the measured power spectral density. Target distance and velocity were computed using the beat frequencies $f_1$ and $f_2$ recorded during the up- and down-chirps respectively. The distance $d$ is given by
\begin{align}
d = \frac{c (f_1 + f_2)}{4r},
\end{align}
and the velocity $v$ is
\begin{align}
v = \frac{\lambda_0 (f_1 - f_2)}{4},
\end{align}
where $r$ is the chirp ramp rate, $c$ is the speed of light, and $\lambda_0$ is the laser wavelength.

\subsection{Optical Efficiency.}
Ideally, the transmitter illumination pattern should exactly match the readout pattern, so that only the receiver pixels currently being read out are illuminated by the transmitter. However, in our current prototype, each transmitter steering position illuminates a field of view of 32 receiver pixels, and 8 receiver pixels are read out at a time. This mismatch was due to a combination of chip area constraints and particularly large $1~\mathrm{mm}$ long thermo-optic phase shifters, and can be resolved by using existing designs for compact thermo-optic shifters \cite{smendezastudillo_oe2019}.

Correcting for the $4\times$ mismatch between the number of transmitter and receiver positions, our equivalent point rate is $2 \times 10^4$ per second. Combined with a $4~\mathrm{mW}$ emitter power, this yields an optical efficiency of $0.2~\mathrm{\mu J/point}$. The combination of high optical efficiency and long range necessary for autonomous navigation applications\cite{curmson_jofr2008} is typically met using mechanically steered LiDARs, such as the commonly used Velodyne VLP-16. This $100~\mathrm{m}$ class mechanical LiDAR uses the same $0.2~\mathrm{\mu J}$ of light per point as our system\cite{jkidd_mscthesis2017}, and has a much poorer depth accuracy of $3~\mathrm{cm}$.

\subsection{Electro-optic characterization.}
A $3.5~\mathrm{GHz}$ oscilloscope (Tektronix DPO7354C) was used for all electro-optic characterization of the receiver array. The amplifier noise floor and shot noise were averaged over a bandwidth of $1 - 3~\mathrm{MHz}$, avoiding low frequency $1/f$ noise from the amplifier, as well as the relative intensity noise peak of the laser. The measured shot noise floor was used to determine the exact local oscillator power at each receiver pixel, since the shot noise power spectral density depends only upon the photocurrent. The common mode rejection ratio of the receiver pixels was measured by modulating the amplitude of the local oscillator light at $10~\mathrm{MHz}$, and comparing the measured electrical output amplitude to the expected amplitude given the local oscillator power and amplifier gain. Based on circuit modelling and independent verification using a test structure on the chip, the total gain of the amplifier chain was $20~\mathrm{k\Omega}$. The total electrical power consumption of the receiver chip was $250~\mathrm{mW}$. This includes all 8 receiver output channels of signal amplification and multiplexing.

\subsection{Characterization of measurement accuracy.}
Measurement error in our 3D imaging system can be divided into two categories: systemic errors due to non-idealities in our system, and random fluctuations in the measured beat frequencies due to shot noise, laser relative intensity noise, laser frequency fluctuations, and electronic noise sources. Systemic errors in our system are very tightly controlled. Since the frequency chirps in our system are generated using direct digital synthesis in an AWG, distance accuracy is fundamentally derived from the speed of light, a fixed physical constant, and the timing accuracy of the clocks in the AWG and ADCs, which are controlled to within a few parts per million. The only remaining source of systemic error comes from optical path length differences between pixels, which manifest as static offsets in measured depth. These are due to differences in on-chip optical waveguide lengths, in addition to subtly differing paths taken through the free space optics by light from different pixels. Since these path length differences are static, they can be eliminated using straightforward calibration measurements.

Thus, the key parameter for our system is depth noise, the variation in depth measurements due to stochastic noise in our system. Depth noise was measured by acquiring 40 sequential frames of a static test target for the histograms shown in Fig. {\ref{fig:4_results}}(b), and 20 sequential frames for the depth precision measurements in Extended Data Fig. {\ref{extfig:5_system_perf}}(b). The mean distance value for each pixel was taken to be the true distance, and depth error was defined as the deviation from the true distance for each pixel. Finally, we defined measurement precision as the standard deviation of the depth error.

\subsection{Contrast measurement.}
Imaging contrast for the system was measured using retroreflective sheeting (white 3M Scotchlite 680CR) as a target. The edge of the retroreflective sheet was placed across the field of view of the system, allowing for the measurement of pixel-to-pixel crosstalk. The edges were deliberately placed near the middle of the $8 \times 8$ pixel receiver blocks to maximize the expected crosstalk. Extended Data Fig. {\ref{extfig:5_system_perf}}(a) illustrates the mean signal strength as a function of distance from the retroreflector edge. This was obtained by averaging the signal strength of 12 pixels in the direction parallel to the retroreflector edge.

The imaging contrast is slightly worse for horizontal edges. This is likely due to the greater numerical aperture of the receiver grating couplers in the vertical direction, leading to additional blurring due to lens aberrations in the vertical direction.

\subsection{Transimpedance amplifier comparison.}
Fig. \ref{fig:3_electrooptic_perf}(d) plots input-referred noise current density against the transimpedance gain-bandwidth product for several state-of-the-art CMOS and BiCMOS optical receiver publications\cite{ahmed_jssc2019, shahdoost_ana2016, mohan_jssc2000, razavi_isscc2000, romanova_elec2019}. A custom design must simultaneously meet requirements for gain, noise, and bandwidth. Generally, the gain-bandwidth product will be constant for a target technology and power consumption. In a resistive shunt-feedback configuration, the input-referred current noise $i_{n,rms}$ is typically dominated by the feedback resistance $R_F$:
\begin{equation}
    i_{n,rms} = \sqrt{\frac{4kT}{R_F} \cdot BW_{-3dB}}
\end{equation}
where $k$ is the Boltzmann constant, $T$ is the temperature, and $BW_{-3dB}$ is the 3-dB bandwidth.
The gain is approximately equal to the feedback resistance, and the bandwidth is determined by the pole at the input, where $C_T$ is the total capacitance at the TIA input, and $A_0$ is the open-loop gain of the TIA.
\begin{equation}
    BW_{-3dB} = \frac{1+A_0}{2\pi \cdot R_F \cdot C_T}
\end{equation}
The negative feedback acts to reduce the input impedance looking into the TIA.
Due to our low bandwidth requirement ($<1~\mathrm{GHz}$), and small diode and parasitic capacitance at the TIA input, we can use a large resistor to get high TIA gain resulting in a reasonable gain-bandwidth product while allowing a low input-referred noise density. Having low-noise electronics improves the system’s detection probability, providing longer range for a given optical power.

\subsection{Pixel Size Reduction.}
When considering the maximum resolution sensor that can fit on a chip with a given area, the pixel size is the dominant factor. The switching trees are negligible in size compared to the receiver FPA if existing compact designs for thermo-optic phase shifters are used. For example, efficient thermo-optic phase shifters as short as $35~\mathrm{\mu m}$ in length have been demonstrated \cite{smendezastudillo_oe2019}.

The current pixel size is limited by the use of foundry PDK devices, which were not designed to minimize footprint. In terms of individual photonic components, efficient grating couplers\cite{jsun_nature2013} with a footprint of $3\times 3~\mathrm{\mu m}^2$, and $2 \times 2$ couplers\cite{aypiggott_acsphoton2020} as small as $3 \times 1~\mathrm{\mu m}^2$ have been demonstrated. Meanwhile, photodiodes with a footprint of $3 \times 1~\mathrm{\mu m}^2$ are feasible due to the short absorption length of germanium\cite{vsorianello_apl2008}. Lastly, using advanced deep sub-micron CMOS technology nodes, the TIA could be reduced to $3\times 3~\mathrm{\mu m}^2$ in size. This should allow the components for a single pixel to fit in an $8\times 5~\mathrm{\mu m}^2$ footprint.

%% Here is the endmatter stuff: Supplementary Info, etc.
%% Use \item's to separate, default label is "Acknowledgements"

%%
%% Extended Data
%%
%% If there are any tables or figures for Extended Data, put them here.
%%
\renewcommand{\figurename}{Extended Data Figure}
\setcounter{figure}{0}    

\begin{figure*}
\centering
\includegraphics[width=0.923\textwidth]{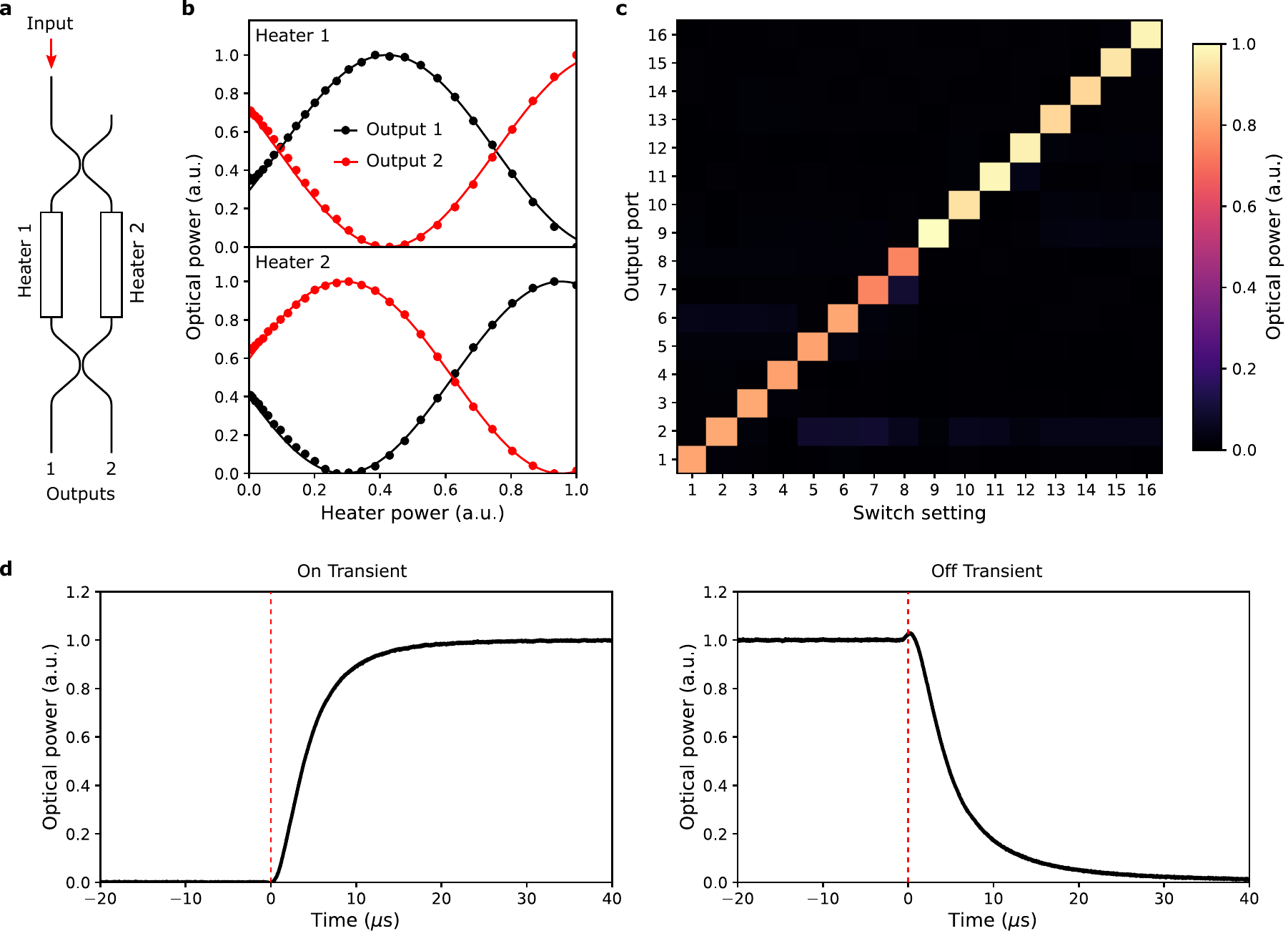}
\caption{Thermo-optic switching tree demonstration. (a) The thermo-optic switches consist of a Mach-Zehnder interferometer with an electrical heater on each arm. (b) Tuning curve for a single thermo-optic switch, showing optical power in the two outputs as a function of applied heater power. The use of two heaters allows the average electrical power consumption per switch to be halved. (c) Output power distribution of the $1 \times 16$ transmitter switch tree for all switch settings, demonstrating clean switching. Output power was monitored using a set of monitor photodiodes at the output of the switch tree. (d) On and off transients for a representative thermo-optic switch, demonstrating $90\% - 10\%$ switching times of $9.1~\mathrm{\mu s}$ and $12.1~\mathrm{\mu s}$ respectively. Due to minor thermal crosstalk between switches, the switching transients are not perfect decaying exponentials.}
\label{extfig:1_switch_trees}
\end{figure*}

\begin{figure*}
\centering
\includegraphics[width=0.95\textwidth]{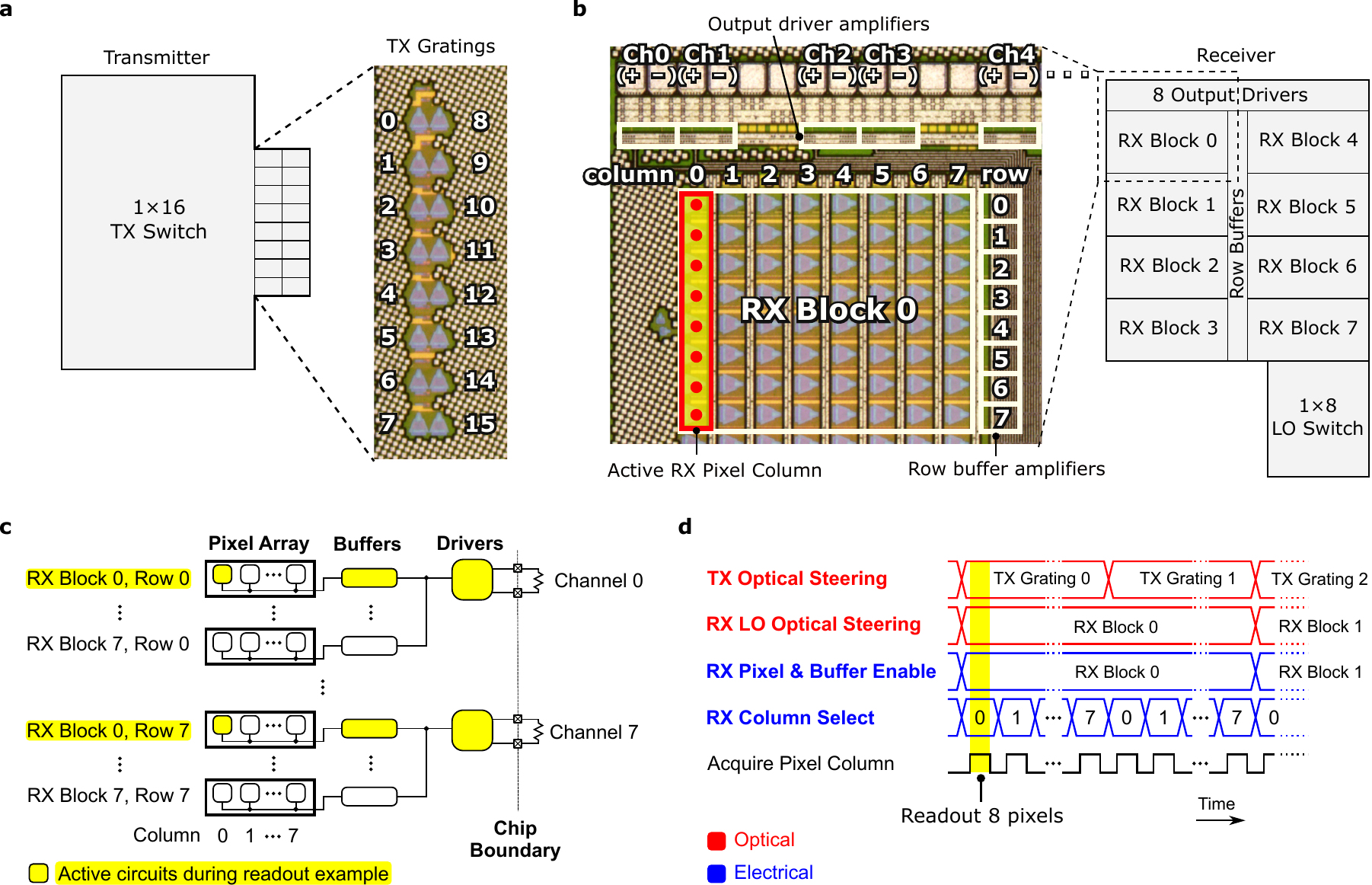}
\caption{Transmitter (TX) and receiver (RX) synchronization and readout architecture. (a) The TX steers light through a 4 level tree of $1 \times 2$ switches to feed the focal plane array of 16 output grating couplers. Each leaf contains a fractional tap and monitor photodiode enabling electronic calibration of the tree. (b) The RX array is divided into 8 blocks of 64 pixels. Imaging an 8-pixel column requires both steering the local oscillator (LO) light to the block and enabling the associated electronics (pixel column and row buffer amplifiers). Signals from the active pixel column are driven by 8 output amplifiers for parallel readout. (c) Several levels of multiplexing are used to map 512 pixels down to 8 output channels. An active RX block has one active pixel per row, with the other disabled pixels within the row presenting high output impedance (no drive strength). The row buffers are similarly passively multiplexed between the blocks. The 8 drivers are always activated during readout. (d) Timing diagram showing synchronization between the optical switching trees (TX and RX) and the electrical readout circuitry.}
\label{extfig:2_readout_arch_and_synch}
\end{figure*}

\begin{figure*}
\centering
\includegraphics[width=\textwidth]{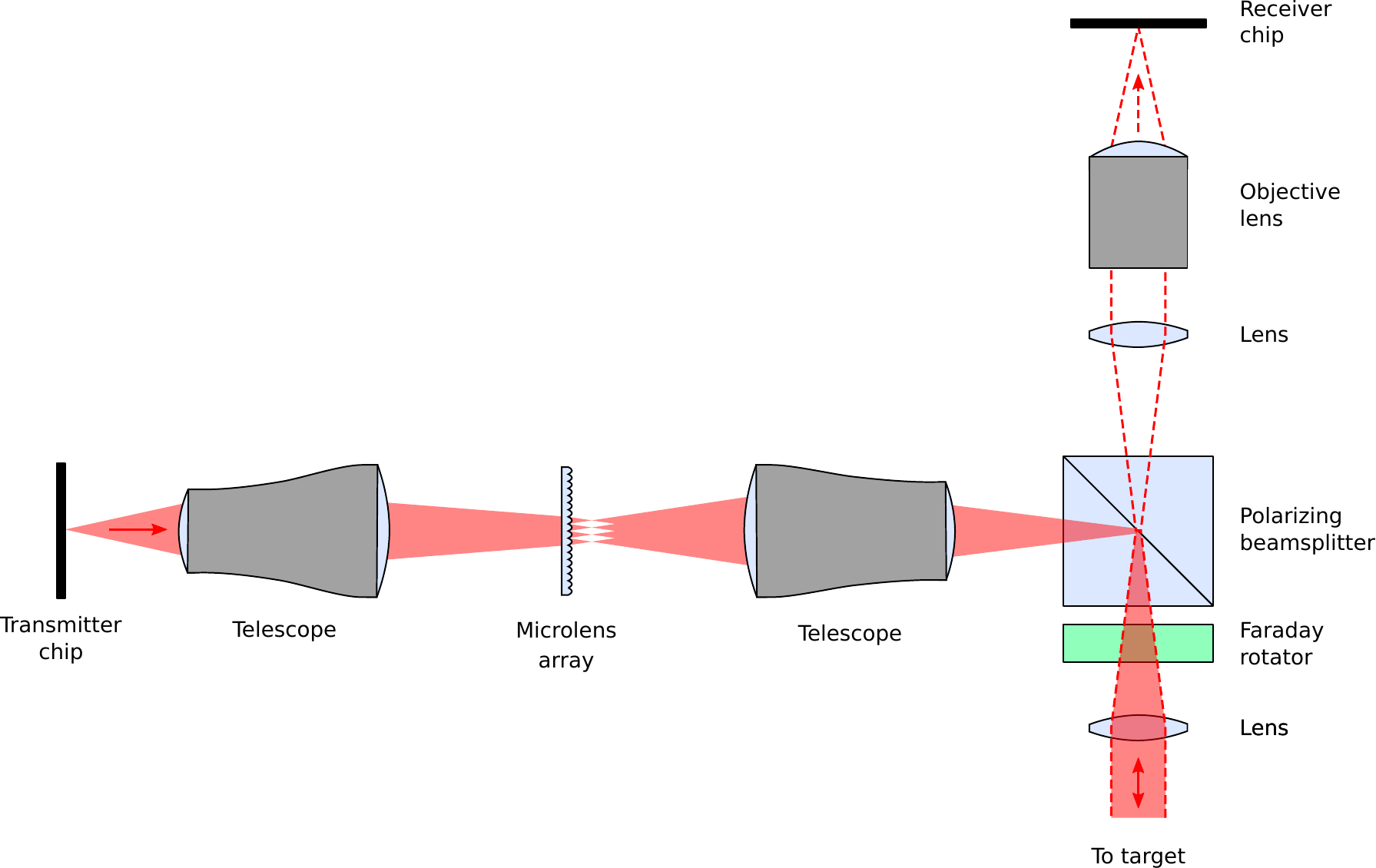}
\caption{Free-space optics schematic of the demonstration system. Much of the complexity in the optical system is to match the receiver and transmitter focal plane arrays, which can be corrected in the future by adjusting the chip layouts. For inexpensive consumer versions of the system, the Faraday rotator and polarizing beamsplitter could be replaced by a 50-50 beamsplitter, at the cost of a $4\times$ reduction in signal strength. Although it is possible to implement this experiment using a single chip for both transmit and receive functions, we have used two identical chips acting as the transmitter and receiver respectively to simplify the experimental setup.}
\label{extfig:3_optical_schematic}
\end{figure*}

\begin{figure*}
\centering
\includegraphics[width=\textwidth]{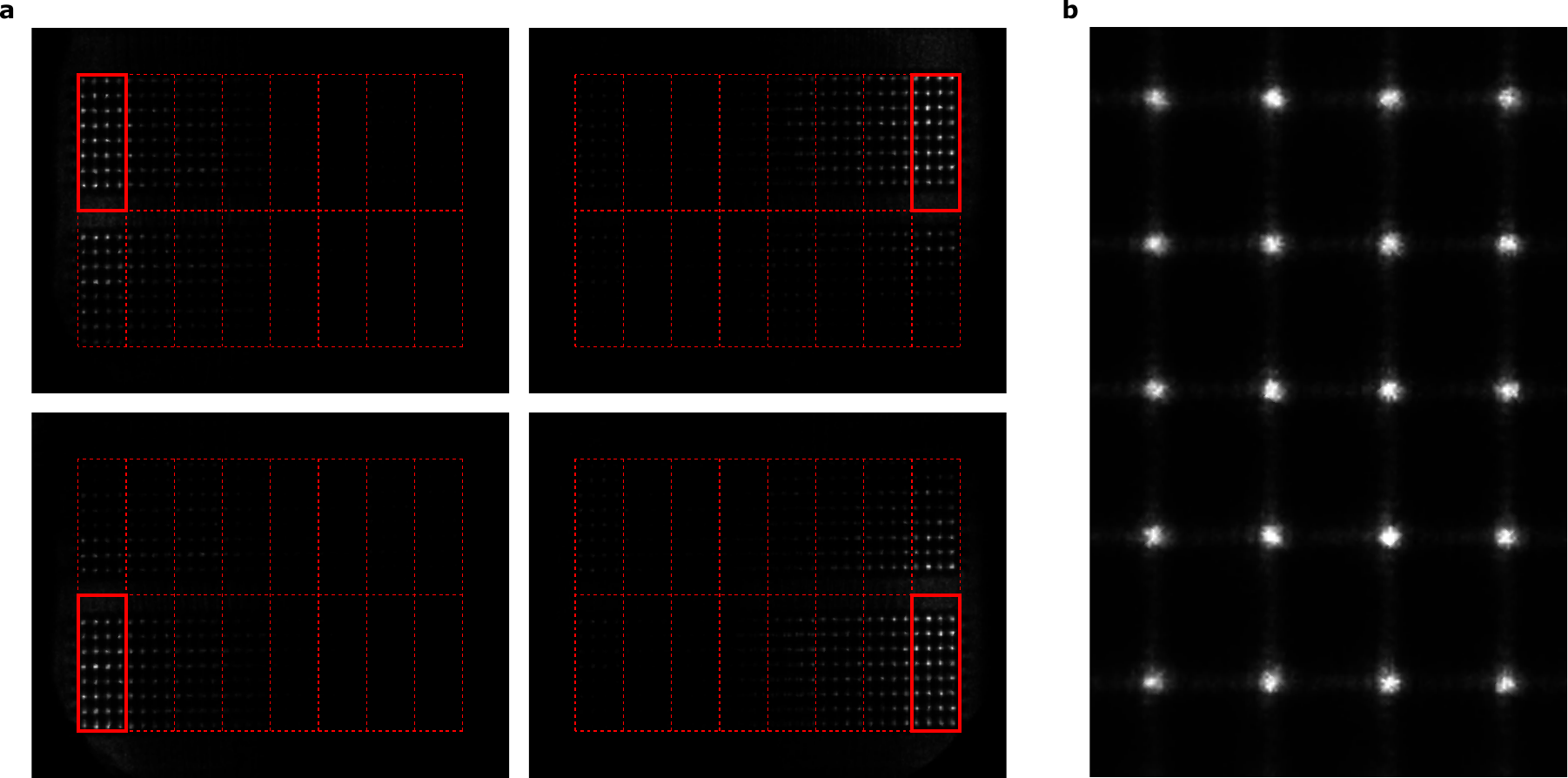}
\caption{Far-field infrared camera images of transmitter steering. (a) Images of several representative steering positions. The receiver fields of view corresponding to the 16 steering positions are indicated by the dashed lines, with the currently active block indicated by a solid outline. The light from the active transmit grating is first slightly defocused to completely illuminate the active block, and then structured by the microlens array. Due to this defocusing and the soft edges of the beam, a small fraction of the transmitted light falls outside of the active block. (b) A zoomed-in image showing the structured illumination pattern produced by the microlens array. The locations of the bright spots coincide with the receiver pixel grating couplers.}
\label{extfig:4_farfield}
\end{figure*}

\begin{figure*}
\centering
\includegraphics[width=0.95\textwidth]{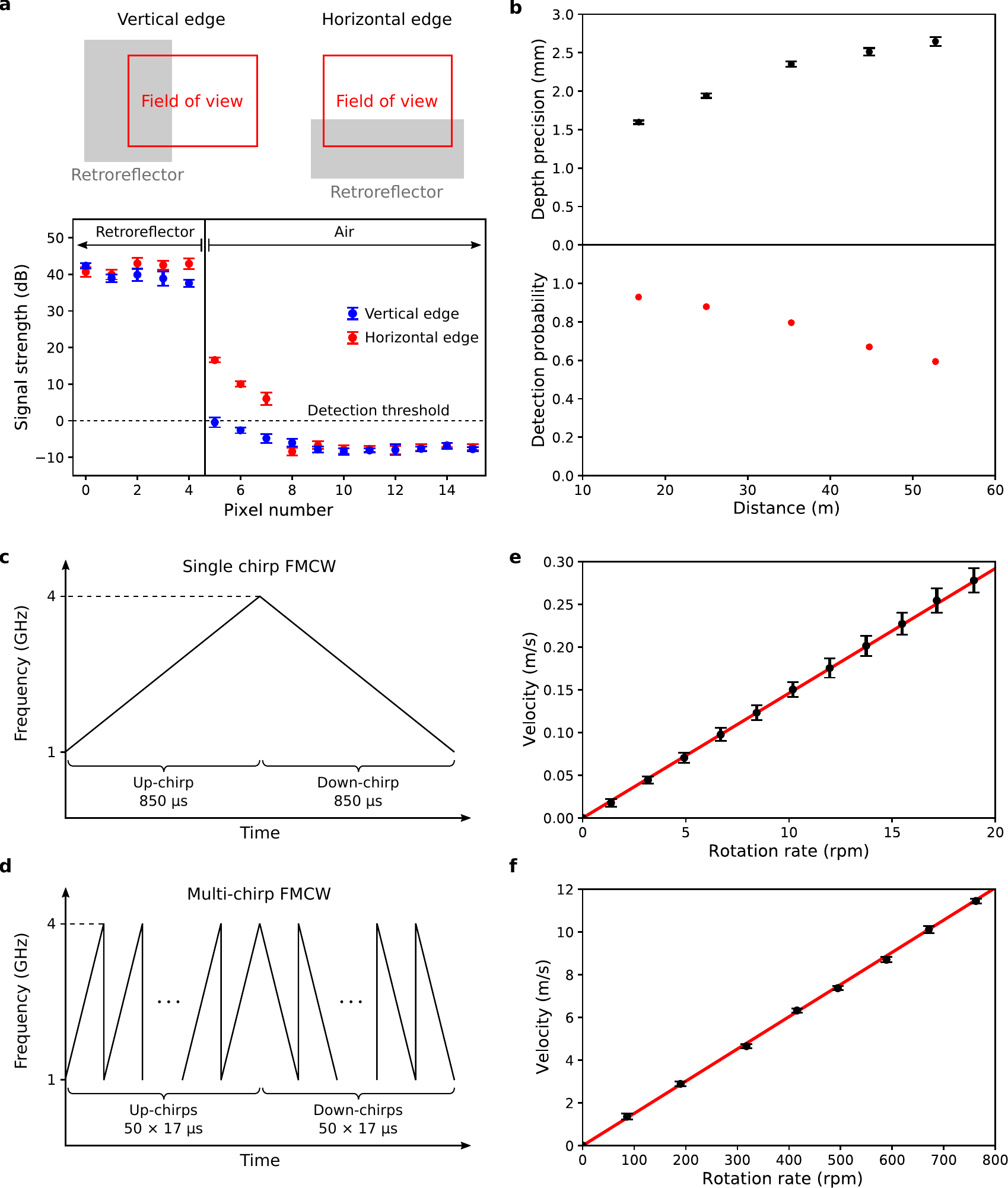}
\caption{Additional characterization of system performance. (a) Imaging contrast measured using retroreflective sheeting. Our system achieves $>25~\mathrm{dB}$ contrast for a 1 pixel displacement, $>50~\mathrm{dB}$ contrast for a 4 pixel displacement, and reaches the system noise floor thereafter, illustrating the excellent pixel-to-pixel isolation in our system. Here, the error bars represent the standard error. (b) Depth precision and detection probability as a function of distance for a $44\%$ reflectance target. The error bars on the depth precision represent the $95\%$ confidence intervals. (c, d) Single-chirp (c) and multi-chirp (d) frequency-modulated continuous-wave waveforms used for measuring slow and fast objects respectively. (e, f) Measured velocity as a function of rotation rate for a $30~\mathrm{cm}$ diameter styrofoam cylinder at a distance of $17~\mathrm{m}$ using (e) single-chirp and (f) multi-chirp waveforms, with error bars indicating the standard deviation.}
\label{extfig:5_system_perf}
\end{figure*}

\end{document}